\documentclass[aps,pra,epsfigure,notitlepage,twocolumn,longbibliography,superscriptaddress]{revtex4-1}

\usepackage{bm} 
\usepackage{graphicx}
\usepackage{amsmath}    
\usepackage{latexsym}
\usepackage{amsfonts}   
\usepackage{amssymb}
\usepackage{comment}
\usepackage{array}      
\usepackage{epsfig}
\usepackage{txfonts}
\usepackage{xcolor}
\usepackage[colorlinks=true,linkcolor=blue,urlcolor=blue,citecolor=blue,pdfusetitle]{hyperref}
\usepackage{hyperref}
\usepackage{ulem}

\newcommand{\ket}[1]{\left|#1\right\rangle}
\newcommand{\bra}[1]{\langle#1|}

\newcommand{\ketbra}[2]{|#1\rangle\langle#2|}
\newcommand{\QD}{Quantum Darwinism}

\newcommand{\nbar}{\bar{n}}

\newcommand{\steve}[1]{{\color{black}#1}}

\begin{document}
\title{Commutativity and the emergence of classical objectivity}
\author{Eoghan Ryan}
\affiliation{Centre for Quantum Materials and Technology, School of Mathematics and Physics, Queen's University, Belfast BT7 1NN, United Kingdom}
\author{Eoin Carolan}
\affiliation{School of Physics, University College Dublin, Belfield, Dublin 4, Ireland}
\affiliation{Centre for Quantum Engineering, Science, and Technology, University College Dublin, Belfield, Dublin 4, Ireland}
\author{Steve Campbell}
\affiliation{School of Physics, University College Dublin, Belfield, Dublin 4, Ireland}
\affiliation{Centre for Quantum Engineering, Science, and Technology, University College Dublin, Belfield, Dublin 4, Ireland}
\author{Mauro Paternostro}
\affiliation{Centre for Quantum Materials and Technology, School of Mathematics and Physics, Queen's University, Belfast BT7 1NN, United Kingdom}

\begin{abstract}
We examine how the ability of a system to redundantly proliferate relevant information about its pointer states is affected when it is coupled to multiple baths. To this end, we consider a system in contact with two baths: one -- termed the {\it accessible} environment  -- which, on its own, induces a pure dephasing mechanism on the state of the system and satisfies the conditions for classical objectivity to be established. The second environment, which we dub as {\it inaccessible}, affects the system in two physically relevant ways. Firstly, we consider an interaction that commutes with the Hamiltonian describing the interaction between system and accessible bath. It thus also gives rise to dephasing of the system, albeit on different time scales. Secondly, we consider a thermalising interaction, which does not commute with the system-accessible environment Hamiltonian. While the former still allows the system to redundantly encode its state into the accessible environment, the latter degrades the correlations, eventually destroying them in the long-time limit, and thus leads to a loss of the conditions necessary for classical objectivity to be established. This sheds light on the role that commutativity between the various system-bath interaction terms plays when establishing the conditions for classical objectivity to be supported.   
\end{abstract}
\date{\today}
\maketitle

\section{Introduction}
Decoherence theory provides the framework to understand the emergence of classical states from an underlying quantum dynamics~\cite{ZurekRMP}. It posits that the nature of the system-environment interaction singles out a set of system states--the pointer states--which form a basis for the system's description and are robust to the deleterious effects of the interaction. It is the commutativity between the system-environment interaction and the pointer basis that determines the (classical) state which the system is driven to by the dynamics~\cite{ZurekRMP}. 

While decoherence accounts for how classicality is achieved, it must be augmented to address the more general question of how we perceive classically objective states~\cite{ZurekNatPhys2009}. This is due to the fact that decoherence simply accounts for the irretrievable loss of coherence due to environmental interactions. Quantum Darwinism~\cite{ZurekNatPhys2009, KohoutPRA2006}, and the more stringent strong quantum Darwinism~\cite{LePRL2018, LePRA2018, Le2020} and spectrum broadcast structures~\cite{HorodeckiPRA2015, KorbiczPRL2014, KorbiczPRA2017}, attempt to address this issue in a mathematically rigorous manner by treating the environment in a more active manner. The core tenet of quantum Darwinism is that for a classically objective state to emerge, the system must proliferate information about its configuration in the pointer basis to the environment. The total environment, $\mathcal{R}$, is therefore considered as a collection of smaller fragments, $\mathcal{F}$, with which the system interacts and is able to share information. The principle quantity of interest is then the quantum mutual information  
\begin{equation}
\label{MutInfo}
\mathcal{I}(S:\mathcal{E}_{f})=H(\rho_S) +H(\rho_{\mathcal{E}_{f}})-H(\rho_S,\rho_{\mathcal{E}_{f}})
\end{equation}
where $H(\cdot)$ denotes the von Neumann entropy, $\rho_S$ is the density matrix of the system, and $\rho_{\mathcal{E}_{f}}$ is the density matrix of the fraction of the environment, $f\!=\!\mathcal{F}/\mathcal{R}$, which an observer has access to. When $\mathcal{I}(S:\mathcal{E}_{f})\!=\!H(S)$, the information about the system is stored completely in the fragment $\mathcal{E}_{f}$  and thus an observer able to interrogate this fragment will have access to all the available system information, and importantly, no additional information can be obtained even if a larger fraction of the environment is accessible~\cite{ZurekNatPhys2009}. Such a condition naturally implies a notion of objectivity, as two observers querying different fragments of the environment will nevertheless have access to the same system information. The system is therefore said to have redundantly encoded its state into the environment degrees of freedom and this redundancy is witnessed by a characteristic plateau in the mutual information, Eq.~\eqref{MutInfo}, for increasingly larger fractions of the environment. 

This framework has been extensively explored for a system in contact with a single, possibly complex, bath~\cite{ZurekSciRep2013, ZurekSciRep2016, ZurekPRA2017, AdessoPRL2018, BalaneskovicaEPJD2015, MendlerEPJD2016, SalvatorePRR, SabrinaPRR, GarrawayPRA2017, ZwolakPRA2010, ZwolakPRL2009, AkramPRL, MirkinEntropy, CakmakEntropy, RyanPLA, ZwolakNJP2012, CampbellPRA2019, LeEntropy, KorbiczPRA19, Korbicz_Quantum, KorbiczPRA2019b, KorbiczPRA2021, DarwinismExp1, DarwinismExp2, DarwinismExp3} where the role of different bath characteristics can have a significant affect on the system's ability to redundantly encode its information within the bath~\cite{LewensteinPRA2017, GalveSciRep2016, GiorgiPRA2015, NadiaPRA, MegierEntropy}. Here, we consider a complementary setting where the system is in contact with two baths, one which we refer to as the ``accessible" environment which consists of the fragments that hypothetical observers would be able to measure. We will assume that this environment gives rise to a purely dephasing dynamics on the system which, in the absence of any other influences, provides the conditions necessary for quantum Darwinism to be exhibited. In addition we assume that the system is also in contact with a second ``inaccessible" bath. By first exploiting a collisional model approach~\cite{CiccarelloReview, SteveEPL, SabrinaNPJQI} we explore how the nature of this inaccessible bath can strongly affect the system's ability to redundantly encode information about its pointer states into the accessible environment's degrees of freedom. We establish that the commutativity between the two Hamiltonians governing how the system interacts with the respective baths dictates whether conditions for classical objectivity can be maintained in the long-time limit. This is a particularly relevant point with strong implications for the ongoing attempts at unveiling any effect that possible non-Markovian features emerging, for instance, from the finiteness of one of the environments in contact with the system would induce~\cite{NadiaPRA,GiorgiPRA2015,GalveSciRep2016, MegierEntropy}. In particular, we show that the emergence of non-Markovian features has no correlation with the success of redundant encoding of information about the system in the state of the accessible environment. 

The remainder of the work is organised as follows. In Sec.~\ref{collision} we examine the microscopic model for the multiple bath setting, considering a minimal model for the accessible environment consisting of three qubits inducing a pure dephasing dynamics on the system, while we employ a collision model to simulate the inaccessible environment. The (rescaled) mutual information between fragments of the accessible environment, labelled $\mathcal{E}_f$, and the system
\begin{equation}
\label{MIeq}
\overline{\mathcal{I}} = \frac{\mathcal{I}(S:\mathcal{E}_f)}{H(S)}
\end{equation}
will be a key figure of merit. Sec.~\ref{commutative} discusses the role of commutativity between the Hamiltonians governing the system-environment interactions and establishes that non-commuting interactions stifle the system's ability to redundantly encode information about its pointer state configuration. Sec.~\ref{nonM} addresses the features the open dynamics of the system from the viewpoint of potential non-Markovian features, showing the lack of correlation between the occurrence of a back-flow of information -- typically considered as one of the underlying fundamental mechanisms giving rise to non-Markovian dynamics -- and the successful redundant encoding of information on the state of the system. Finally in Sec.~\ref{conc} we draw our conclusions. 

\section{Collisional-model picture of the system-environment interaction}
\label{collision}
\begin{figure*}
{\bf (a)} \hskip0.3\linewidth {\bf (b)} \hskip0.34\linewidth {\bf (c)}\\
\includegraphics[width=0.25\linewidth]{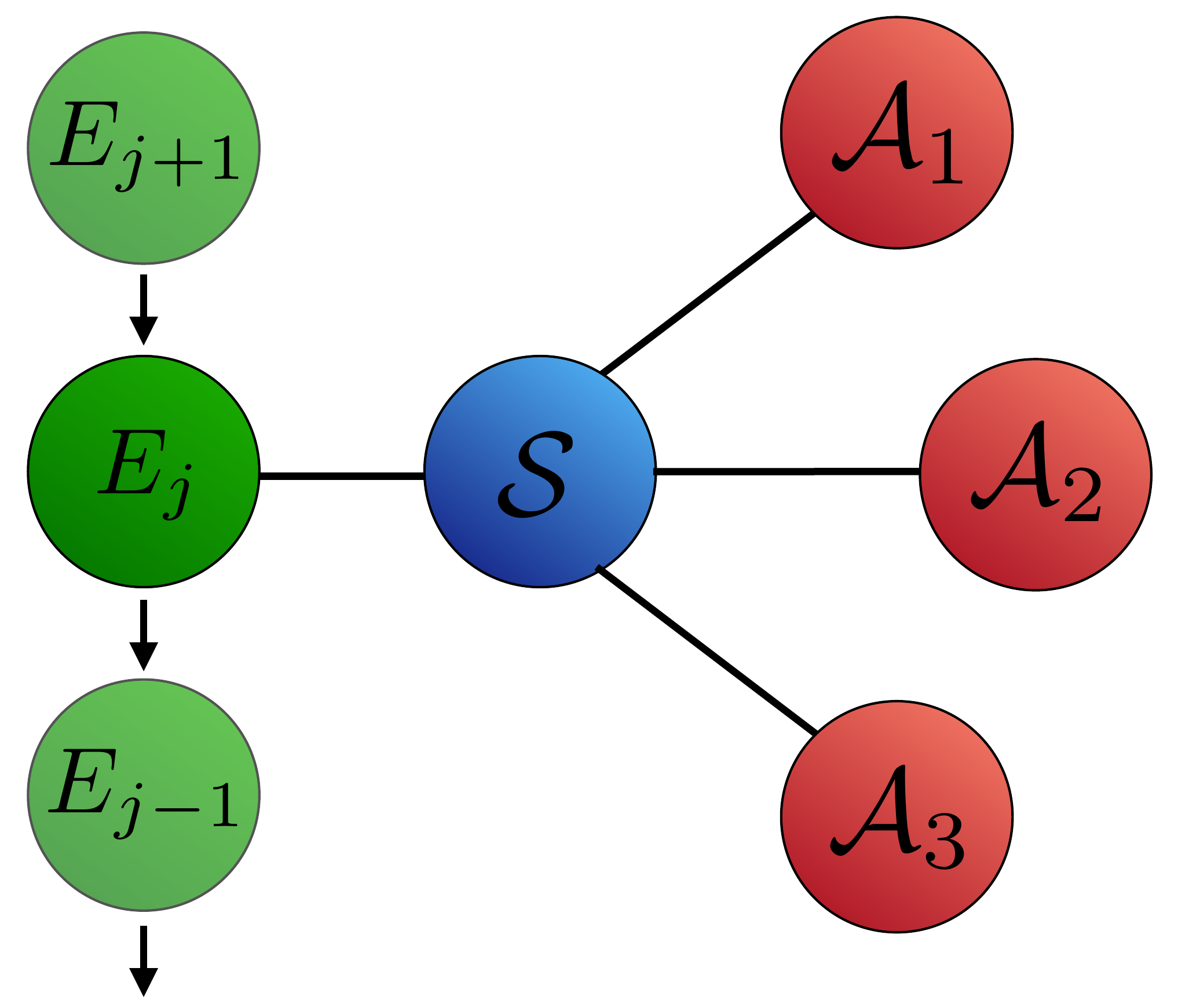}~~~~\includegraphics[width=0.33\linewidth]{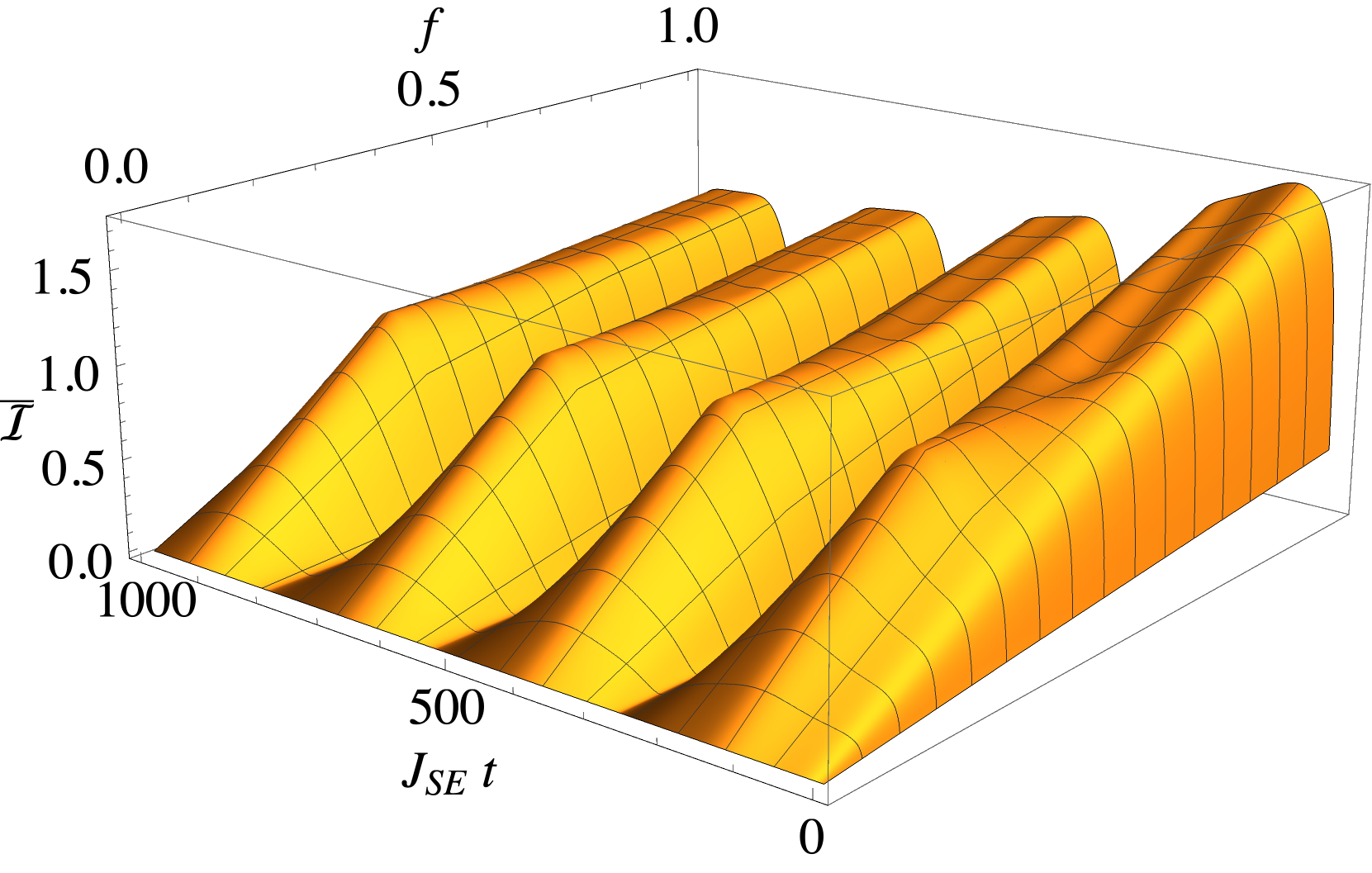}~~~~\includegraphics[width=0.33\linewidth]{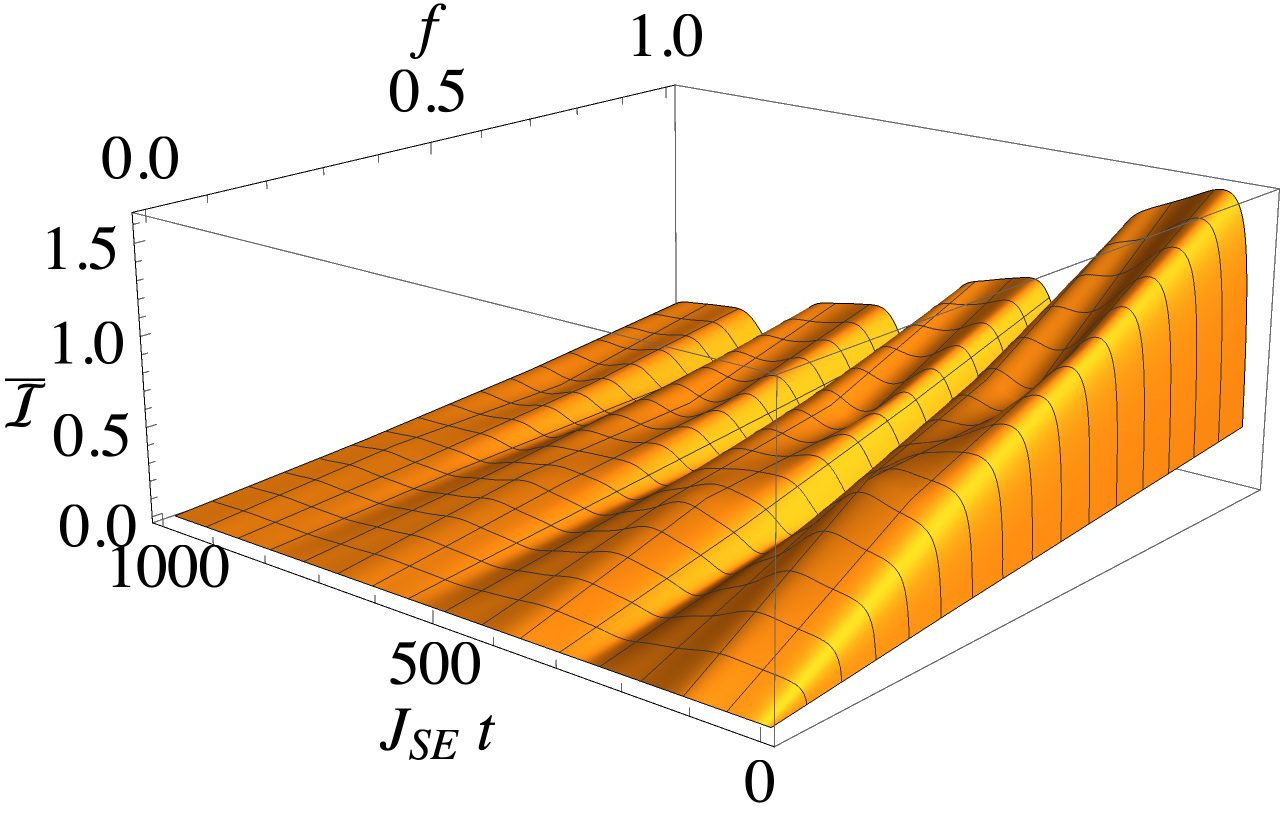}
\caption{{\bf (a)} Schematic of the microscopic collision model employed. {\bf (b,c)} Rescaled mutual information, $\overline{\mathcal{I}}$, for {\bf (b)} dephasing collisional environment and {\bf (c)} thermalising collisional environment. We fix $J_{SA} \tau_1=0.0075\pi/4$ and $J_{SE} \tau_2=0.015\pi/2$ and $\beta\!=\!0$. For both interactions the collisional environment drives the system towards a \steve{fully decohered} state, however, in the case of a dephasing interaction the mutual information shared between system and accessible environment fragments shows the characteristic redundancy plateau transiently emerges with period dictated by the $J_{SA}$ interaction strength. For a thermalising collisional bath there is an overall envelope where the redundancy plateaux are progressively damped. \steve{As elucidated in Sec.~\ref{commutative} this behaviour is explained by the commutativity, or lack thereof, between the system-environment interaction and system-accessible fragment interaction. In the dephasing case, the ability of the environmental fragments to create classical correlations with the system is unaffected by the presence of the collisional environment as the interactions commute, while this is not the case for a thermalising environment, which serves to fully decohere the fragments as well as the system in the long time limit. }}
\label{fig1_cm}
\end{figure*}

We consider the situation as depicted in Fig.~\ref{fig1_cm}(a) where the system of interest, $S$ with free Hamiltonian $H_S\!=\!\sigma_z^S$, is in contact with two distinct environments, one composed of a small number of constitutions which we refer to as the accessible environment and represents the degrees of freedom which an observer would have access to. For interactions which give rise to a pure decoherence for the system, i.e. those interactions which only affect the coherences and leave the populations unchanged, it is known that such interactions lead to the type of global system-environment configurations that support classically objective states~\cite{KorbiczPRA19, CampbellPRA2019}. The Hamiltonian governing the interaction between the system of interest and the fragments of the accessible environment, labelled $A_i$, is then
\begin{equation}
\label{HSA}
H_{SA} = J_{SA} \sum_{i} ( \sigma_z^S \otimes \sigma_z^{A_i} ).
\end{equation}
Here $\sigma^k_{p}$ is the $p=x,y,z$ Pauli operator of either the system (for $k=S$) or one of the fragments (when taking $k=A_i$, $\forall i$).
For clarity, in what follows we will restrict the size of the accessible environment to three subsystems, which is the smallest size required for characteristic redundancy plateaux to be observed~\cite{RyanPLA}, however remark that due to the nature of the considered interactions in Eq.~\eqref{HSA} our results remain qualitatively unaffected for larger accessible environments, {with larger accessible environments serving to increase the rate at which redundant encoding occurs.} In the absence of any other influences, this setting is well known to recover clear Darwinistic features. In particular, when system and all environment fragments are qubits, as will be considered throughout the present work, it satisfies the conditions for strong quantum Darwinism, or equivalently, spectrum broadcast structures~\cite{CampbellPRA2019}. 

In addition to the accessible environment, we allow the system of interest to be coupled to a second inaccessible bath, which could in principle be of a different nature. Such a setting is physically well motivated: nothing precludes augmenting the original paradigm of quantum Darwinism to allow for the system to be simultaneously coupled to a thermal bath for instance. Recently the delicate interplay between whether it is possible for states to be both thermal and classically objective has been explored~\cite{LeEntropy}. Here, we address a complementary setting in order to gain qualitative insight into how the nature of the interactions between a system and an inaccessible environment affect the system's ability to redundantly proliferate information into accessible environmental degrees of freedom. To this end, we rely on a collision model description of the inaccessible environment. 

Collision models provide a versatile tool for modelling open system dynamics and are particularly suited to our purposes~\cite{CiccarelloReview, SteveEPL, CampbellPRA2019}; the system interacts with a single incoming unit for a short period of time, after which this unit is traced out and a ``fresh" unit is introduced, thus capturing the inaccessible nature of the environment we are modelling as any information regarding the state of the system which is imprinted on these units is irretrievably lost. A further advantage of exploiting the collisional model framework is that it allows to simulate different physically relevant environmental dynamics by simply tuning the microscopic details of the interaction. Care must be taken in this context: if the system-environment interaction does not commute with the system's Hamiltonian, the free-evolution must be taken into account~\cite{Lorenzo2017, GiacomoPLA}. However, in what follows we consider two system-environment interactions that give rise to physically relevant dynamics, namely dephasing and thermalisation, and are unaffected by the inclusion of the free evolution term. The respective Hamiltonians are
\begin{equation}
\label{SEinteractions}
\begin{aligned}
H_{S{E_j}}^D &= J_{SE} ( \sigma_z^S \otimes \sigma_z^{E_j}),\\ 
H_{S{E_j}}^T &= J_{SE} ( \sigma_x^S \otimes \sigma_x^{E_j} + \sigma_y^S \otimes \sigma_y^{E_j} ),
\end{aligned}
\end{equation}
where $E_j$ is the label for the $j^\text{th}$ unit of the inaccessible environment modelled through the collisional picture.  

The system then interacts stroboscopically with the environments, first colliding for a time $\tau_1$ with the all accessible fragments, then interacting with the collisional bath for a time $\tau_2$, i.e.
\begin{equation}
\rho(n+1) = U_{SE} U_{SA} \rho(n) U_{SA}^\dagger U_{SE}^{\dagger}
\end{equation}
where $U_{SA} = \exp{(-i H_{SA}} \tau_1)$ and $U_{SE} = \exp({-i H_{S{E_j}}} \tau_2)$. The accessible fragments and system are assumed to be initially prepared in state $\ket{+}=(\ket{0}+\ket{1})/\sqrt2$ with $\sigma_z\ket{0}=\ket{0}$ and $\sigma_z\ket{1}=-\ket{1}$. Each incoming collisional unit is initialised in a Gibbs state with dimensionless inverse temperature $\beta$, i.e.
\begin{equation}
    \rho_{E_i}=\frac{1}{2}\begin{pmatrix}
1+\text{tanh}(\beta) & 0 \\
0 & 1-\text{tanh}(\beta)
\end{pmatrix}.
\end{equation}
The system-accessible environment starts in a product state $\rho(0)=\rho_S(0)\bigotimes_{i=1}^{3}\rho_{A_i}(0)$. For simplicity we take infinite temperature collisional units, however we remark, up to some minor qualitative differences, our results hold for finite temperatures. The infinite temperature assumption together with the considered initial states means that the resulting dynamics for the system is identical regardless of whether the interaction between system and collisional bath gives rise to dephasing or thermalisation.

\begin{figure}[b]
{\bf (a)}
\\
\includegraphics[width=\columnwidth]{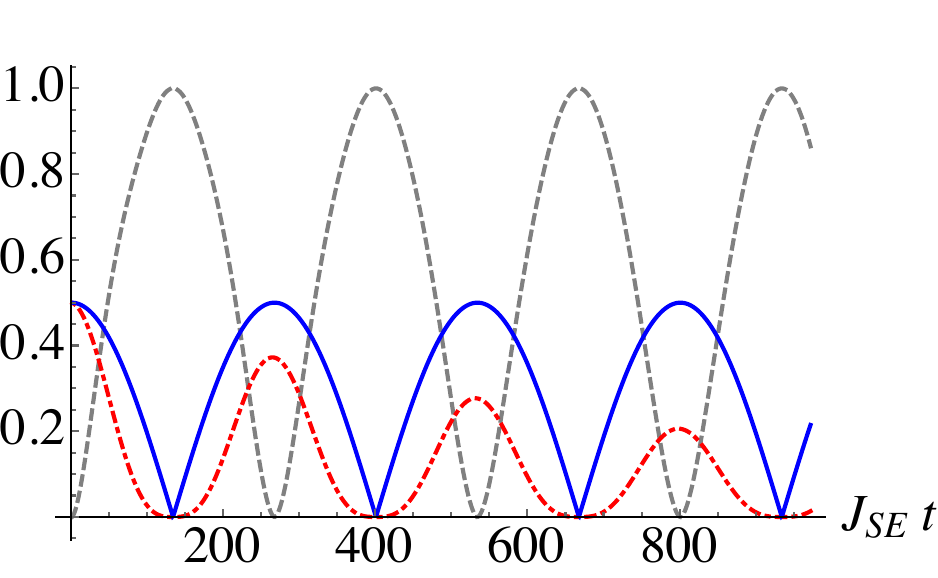}\\
{\bf (b)}\\
\includegraphics[width=\columnwidth]{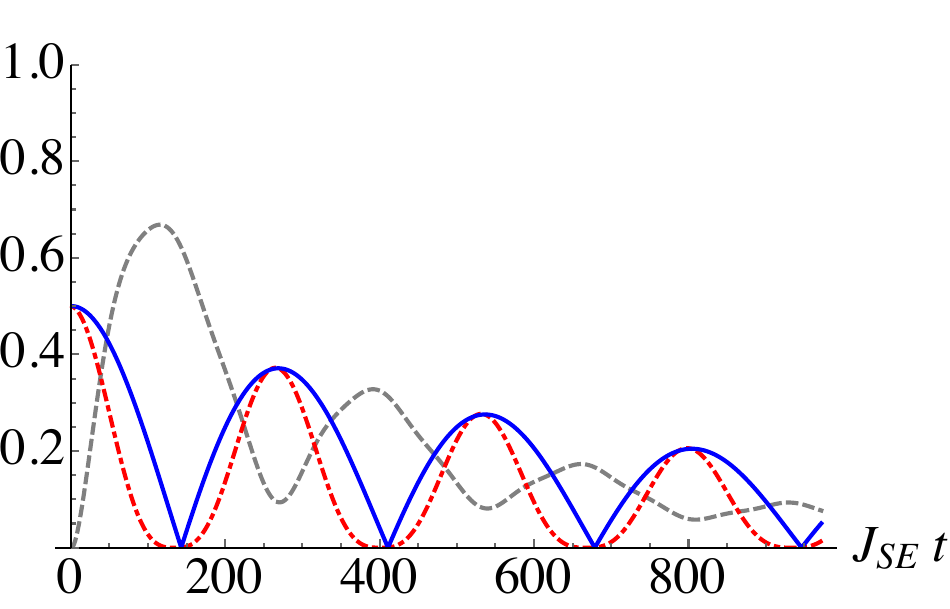}
\caption{Mutual information between the system and a single accessible environmental fragment (dashed grey), the system coherence (dot-dashed red), and the fragment coherence (solid blue). {\bf (a)} Dephasing collisional bath and {\bf (b)} thermalising collisional bath.}
\label{fig_cm2}
\end{figure}
We begin examining the mutual information shared between the accessible fragments and the system. For a dephasing interaction between the collisional bath and the system, Fig.~\ref{fig1_cm}(b) shows that the characteristic redundancy plateau emerges. For short time dynamics, where the effects of the collisional bath are small, the rise in mutual information when an observer has access to all constituents of the accessible environment, and therefore $f\!=\!1$, is evident. For longer times where the collisional bath is able to decohere the system, we find that redundancy of the information shared between the system and accessible fragments is maintained, however, now there is no rise when an observer has access to all accessible fragments. In essence, while the system shares correlations with the collisional bath, it is still able to share the relevant classical information with fragments of the accessible environment. We see that the emergence of the characteristic redundancy plateaux are periodic. This is due to the small size of the considered accessible fragments, with the period being fully determined by the strength of the system-fragment interaction, $J_{SA}\tau_1$. The situation is markedly different when the interaction between system and collisional bath gives rise to thermalisation, as shown in  Fig.~\ref{fig1_cm}(c). The short time dynamics is qualitatively identical, and this is again due to the fact that the timescales are too short for the collisional bath effects to be significant. For longer interaction times, while we see that the mutual information continues to plateau when an observer has access to larger fractions of the accessible environment, its value is no longer equal to the system entropy, thus indicating that while information is being redundantly encoded, an observer cannot gain access to all the classical information about the state of the system due to the interaction with the thermalising environment. For long interaction times we see that the information shared between system and accessible fragments $\overline{\mathcal{I}}\!\to\!0$, indicating the complete loss of all correlations. 

We reiterate that the system undergoes an identical evolution in both cases for the considered parameters. Given the choice of initial conditions, the populations remain fixed and the effect of the collisional bath is to simply dampen the coherences present in the system. We demonstrate this explicitly in Fig.~\ref{fig_cm2} where the dashed red curve shows the behavior of the coherence term of $\rho_S$~\cite{Baumgratz}
\begin{equation}
{\cal C}=\sum_{i,j}\left|\left(\rho_S\right)_{ij}\right|
\end{equation}
and is identical for both dephasing interactions [cf. panel (a)], and thermalising interactions [panel (b)], with the collisional bath. If we focus on only a single accessible fragment, we can see a striking difference that the form of the system-inaccessible bath interaction has on the properties and correlations that the accessible fragment shares with the system. For a dephasing collisional bath, Fig. \ref{fig_cm2}(a), we show the dynamics of the fragment coherence and the (rescaled) mutual information shared between this fragment and the system, solid blue and dashed grey curves, respectively. The clean periodic behavior is exhibited, with perfect classical correlations established between the system and fragment when the accessible qubit's coherence term vanishes, with this behavior persisting regardless of the fact that the system coherence term is being damped by the collisional bath. In contrast, for the thermalising inaccessible bath we find that the accessible fragment losses coherence in line with the behavior of the system. This in turn reduces correlations shared between the system and the accessible fragment, with the overall effect being that all coherence and correlations are destroyed by the thermalising interaction in the long time limit~\cite{LeEntropy}.

Therefore, despite the dynamics of the system being identical in these two situations, the ability for the system to establish the requisite correlations is strongly dependent on the nature of the interaction with the inaccessible environment. While other choices of initial states and/or inaccessible bath temperatures will present minor quantitative differences, the overall picture remains the same: dephasing interactions between the inaccessible bath and the system always allows for perfect redundancy of information to be encoded within the accessible environment, while thermalising interactions will destroy all correlations leading to a loss of classical objectivity. The reason for this difference in behavior is rooted in the non-commutativity between the interactions of the system with the two separate baths, as discussed in the proceeding section.  


\section{Role of commutativity}
\label{commutative}
The previous section demonstrated that the nature of the system-inaccessible environment can have a significant effect on whether the conditions for classical objectivity are fulfilled or not. We can gain insight into the reason for this dichotomy by examining the interplay between the various interactions. \steve{Specifically, we focus on the commutativity -- or lack thereof -- of the Hamiltonians dictating how the system interacts with the two environments.} As the accessible environment fragments are non-interacting, and due to the considered form of the system-fragment Hamiltonians, we can restrict our attention to only a single qubit of the accessible environment since, given the symmetry of setting, the exhibited behavior is identical for each individual accessible fragment. We can recast the problem using standard tools from open quantum systems, with the system simultaneously coupled to a Lindblad bath and a single auxiliary qubit~\cite{CampbellPRA2010, CampbellPRA2012, MariaPRA, SabrinaPRA2018}. The dynamics of the system-accessible fragment is therefore governed by the Markovian master equation
\begin{equation}
\dot{\rho}_{SA}=-i[H_{SA},\rho_{SA}]+\mathcal{L}(\rho_{SA}).
\label{mastereq}
\end{equation}
The superoperator, $\mathcal{L}(\cdot)$, determines the effect that the inaccessible environment has on the system. As previously, we consider both situations where $\mathcal{L}$ gives rise to dephasing and thermalisation affecting the system only, i.e.
\begin{equation}
\begin{aligned}
 \mathcal{L}^D(\rho)&=\gamma(\sigma_z^S\rho\sigma_z^S-\rho),\\
 \mathcal{L}^T(\rho)&=\gamma(\nbar+1)\left(\sigma_-^S\rho\sigma_+^S-\frac{1}{2}\left[\rho,\sigma_+^S\sigma_-^S\right]\right)\\
 &+\gamma \nbar\left(\sigma_+^S\rho\sigma_-^S-\frac{1}{2}\left[\rho,\sigma_-^S\sigma_+^S\right]\right),
\end{aligned}
\end{equation}
where $\nbar=1/(e^\beta-1)$ is the mean number of thermal excitations in the environment. 
We can readily solve Eq.~\eqref{mastereq} for both types of bath and determine the reduced states for both system and accessible fragment. For clarity we fix both the accessible qubit and system to have the same initial state $\rho_S(0)\!=\!\rho_A(0)\!=\! \ketbra{+}{+}$, although remark that our results are qualitatively unaffected for other suitable choices. In the case of dephasing, the system's populations are unaffected, while for a thermalising bath the populations are driven to the relevant values as dictated by the canonical Gibbs state given by the choice of $n$. Regardless of the nature of the system-inaccessible environment interaction, the accessible fragment's populations are invariant. We find that it is the behavior of the coherence terms in the various reduced density matrices that determine whether classically objective states can be achieved. The coherences are given by
\begin{equation}
    \begin{aligned}
&\text{dephasing:}\begin{cases}
\bra{0}\rho_S^{D} \ket{1} = e^{-2\gamma t}\cos\left( J_z t \right),\\
\bra{0}\rho_A^{D} \ket{1} = \cos\left( J_z t \right)/2,
\end{cases}\\
&\text{thermalising:}\begin{cases}
\bra{0}\rho_S^{T} \ket{1}= e^{-\gamma \left(\nbar+\frac{1}{2}\right)t}\cos\left( J_z t \right), \\
\bra{0}\rho_A^{T} \ket{1}=\dfrac{e^{{-(\gamma+2\nbar\gamma+\alpha)t/2}}}{4\alpha}\left[ (e^{\alpha t}{+}1) \alpha{+} \gamma_{\nbar}(e^{\alpha t}{-}1)\right],
\end{cases}
\end{aligned}
\end{equation}
where $\gamma_{\nbar}=\gamma(2\nbar+1)$ and $\alpha\!=\!\sqrt{-4J_z(J_z+i\gamma)+\gamma^2_{\nbar}}$. We see that for both types of environment, the system coherence undergoes two competing effects. The dephasing interaction with the accessible fragment gives rise to an oscillatory behavior, while the interaction with the inaccessible environment leads to an exponential decay. Thus, regardless of the nature of the inaccessible bath, the system coherence will vanish asymptotically. If we turn our attention to the behavior of the fragment's coherence we see the markedly different effect that the nature of the inaccessible environment now has. For a dephasing Lindblad bath, due to the fact that the interactions of the system with the two baths commute, we find that the accessible fragment is blind to the presence of the inaccessible bath, with its coherence term oscillating with a period dictated by the strength of the system-fragment interaction~\cite{CampbellPRA2019}, while the Lindblad bath has no effect on its dynamics. In contrast, for a thermalising bath, we see that despite the accessible fragment not interacting directly with the Lindblad bath, its coherence term is nevertheless exponentially suppressed. This, in conjunction with the behavior of the system coherence term, means that no correlations can be maintained in the long time limit and, thus, leads to a complete loss of the conditions necessary to support classically objective states. It is worth stressing the difference between the system-inaccessible environment interactions, which is clearly seen when considering the microscopic description captured by Eq.~\eqref{SEinteractions}. In the case of dephasing, purely informational exchange occurs, while the thermalising interaction supports both information and energy exchanges, with the latter destroying all correlations in the long time limit.

\section{Analysis of non-Markovian features}
\label{nonM}
In order to investigate the effect the back-flow of information has on the emergence of \QD, we  consider a model for time-dependent dephasing on system $S$ described by the master equation
\begin{equation}
\label{nmDephasing}
\dot{\rho}_{SA}=-i[\rho_{SA},H_{SA}]+(g(t)+\gamma)(\sigma_{z}^{S}\rho_{SA}(t)\sigma_{z}^{S}-\rho_{SA}(t))
\end{equation}
with $H_{SA}$ as in Eq.~\eqref{HSA}, $\gamma$ the standard (constant) rate of Markovian dephasing and $g(t)$ a time-dependent rate that, following the model presented in Ref.~\cite{Souza13}, we take of the form $g(t)=\tan({\cal J} t)/2$ with ${\cal J}$ a typical rate that, for the sake of simplicity, we take as ${\cal J}=J_{SA}$. At $t=\pi/(2J_{SA})$, the time-dependent rate $g(t)$ switches sign, taking negative values until $t=\pi/J_{SA}$. Under the assumption of $|g(t)|\gg\gamma$, this results in the breaking of the divisibility of the underlying dynamical map of the the $S$-$A$ compound, and thus determines the emergence of non-Markovian features~\cite{Rivas2010}. 

\begin{figure}[t]
{\bf (a)} \hskip0.5\columnwidth {\bf (b)}\\
\includegraphics[width=\columnwidth]{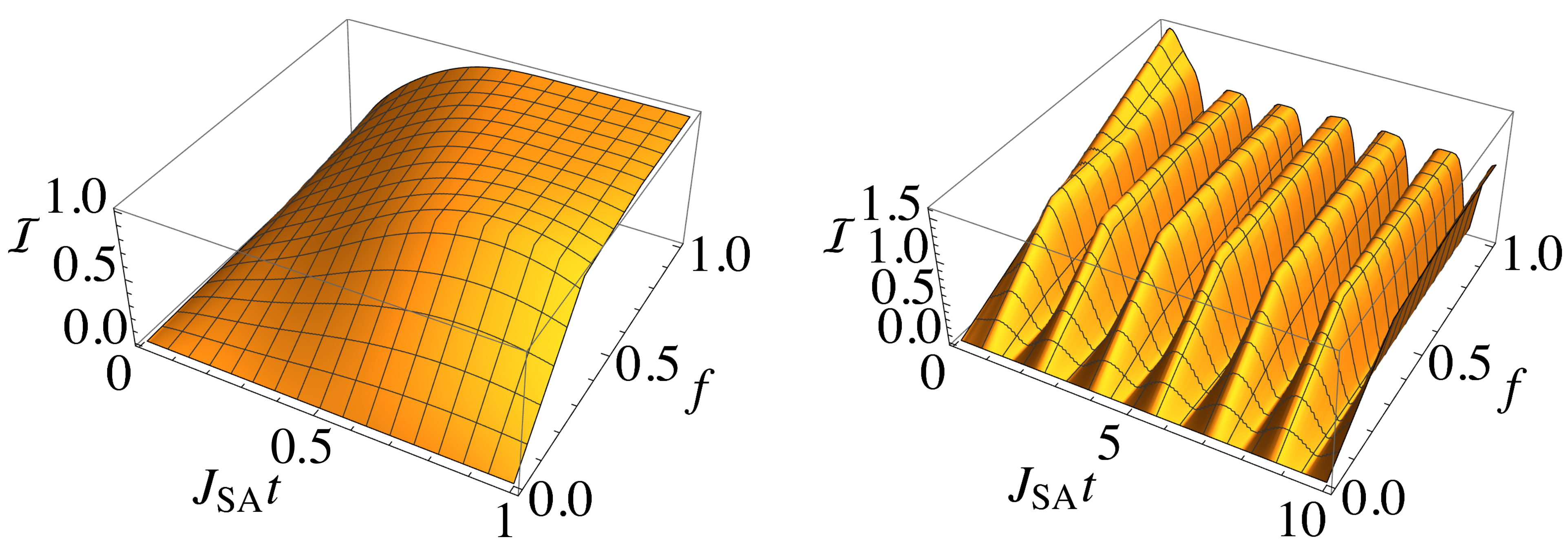}
\caption{Mutual information between system and the accessible environment following the model in Eq.~\eqref{nmDephasing}. In panel {\bf (a)} we used  $J_{SA}/\gamma=1$, resulting in a fully Markovian dynamics within the time-frame considered. In panel {\bf (b)}, we have the strong-coupling condition  $J_{SA}=10\gamma$ and a time-frame within which the rate at which the dephasing mechanism occurs switches sign, thus breaking the divisibility of the underlying dynamics and results in the emergence of non-Markovian effects.}  
\label{fig_non-Mark}
\end{figure}
In Fig~\ref{fig_non-Mark} we report the mutual information between system and accessible environment obtained when taking different values of the $S$-$A$ coupling strength. In Fig.~\ref{fig_non-Mark} {\bf (a)} we consider the case of $J_{SA}/\gamma=1$ that, within the time-window considered, results in a fully Markovian process. 
Fig.~\ref{fig_non-Mark} {\bf (b)} shows instead a situation where $J_{SA}/\gamma=10$, allowing for the rate of dephasing to switch sign and thus give break divisibility already within the considered time window. Remarkably, the behaviors shown in Fig.~\ref{fig_non-Mark} are qualitatively and quantitatively analogous to those obtained in a fully Markovian dephasing model where $g(t)=0$ (with $\gamma\neq0$).  
This hints towards the fundamental irrelevance of a non-Markovian mechanism in the settling of a process of redundant encoding of information. \steve{When considered along the lines of recent literature reporting how non-Markovian dynamics might aid, rather than hinder, the emergence of \QD~~\cite{Oliveira19,LewensteinPRA2017}, our results reinforce the need for a systematic assessment of the role of non-Markovianity in the emergence of objectivity.}



\section{Conclusions}
\label{conc}
We have examined the emergence and suppression of signatures of classical objectivity when a system is in contact with multiple environments. Assuming an observer can query an accessible environment, which interacts with the system via pure dephasing interactions, which are known to support the conditions necessary for the establishment of classically objective states, we have demonstrated that the nature of the interaction of the system with the remaining inaccessible environment(s) can drastically affect the establishment of classical objectivity, as captured by quantum Darwinism. We have shown that for system-inaccessible environment interactions that commute with the system-accessible environment interaction, the relevant system information can proliferate into the accessible fragments since these environmental degrees of freedom are unaffected by the presence of the other bath. However, for other interaction terms which do not commute with the system-accessible Hamiltonian, while partial redundancy can be established transiently, the interaction of the system with the inaccessible environment leads to the loss of all correlations, and thus also of the loss of the conditions for classical objectivity. 

Qualitatively the same behavior is exhibited if the effects due to the inaccessible environment are modelled through master equations giving rise to non-Markovian dynamics. Thus, we establish that rather than the Markovian vs. non-Markovian nature of the bath affecting the system's ability to redundantly encode information in the accessible environment~\cite{GalveSciRep2016, LewensteinPRA2017, GiorgiPRA2015, NadiaPRA}, it is the commutativity of the interactions that dictate whether the conditions for objectivity are met. Our results indicate that commutativity plays a central role in a system's ability to redundantly encode its information and may be complementary to the subtle role that commutativity plays in, for instance, equilibration which has recently been established~\cite{Nicole1, Nicole2, Nicole3} or accurately modelling open system dynamics~\cite{Kolodynski18}.

\acknowledgements
This work is supported by the Northern Ireland Department for Economy (DfE), the Thomas Preston Scholarship,  Irish Research Council Project ID GOIPG/2020/356, the Science Foundation Ireland Starting Investigator Research Grant ``SpeedDemon" No. 18/SIRG/5508, the H2020-FETOPEN-2018-2020 project TEQ (grant No. 766900), the DfE-Science Foundation Ireland (SFI) Investigator Programme (grant 15/IA/2864), the Royal Society Wolfson Research Fellowship (RSWF$\backslash$R3$\backslash$183013), the Royal Society International Exchanges Programme (IEC$\backslash$R2$\backslash$192220), the Leverhulme Trust Research Project Grant (grant No. RGP-2018-266), the UK EPSRC (grant EP/T028424/1), and the Department for the Economy Northern Ireland under the US-Ireland R\&D Partnership Programme

\bibliography{qd_multiE}

\begin{thebibliography}{54}%
\makeatletter
\providecommand \@ifxundefined [1]{%
 \@ifx{#1\undefined}
}%
\providecommand \@ifnum [1]{%
 \ifnum #1\expandafter \@firstoftwo
 \else \expandafter \@secondoftwo
 \fi
}%
\providecommand \@ifx [1]{%
 \ifx #1\expandafter \@firstoftwo
 \else \expandafter \@secondoftwo
 \fi
}%
\providecommand \natexlab [1]{#1}%
\providecommand \enquote  [1]{``#1''}%
\providecommand \bibnamefont  [1]{#1}%
\providecommand \bibfnamefont [1]{#1}%
\providecommand \citenamefont [1]{#1}%
\providecommand \href@noop [0]{\@secondoftwo}%
\providecommand \href [0]{\begingroup \@sanitize@url \@href}%
\providecommand \@href[1]{\@@startlink{#1}\@@href}%
\providecommand \@@href[1]{\endgroup#1\@@endlink}%
\providecommand \@sanitize@url [0]{\catcode `\\12\catcode `\$12\catcode
  `\&12\catcode `\#12\catcode `\^12\catcode `\_12\catcode `\%12\relax}%
\providecommand \@@startlink[1]{}%
\providecommand \@@endlink[0]{}%
\providecommand \url  [0]{\begingroup\@sanitize@url \@url }%
\providecommand \@url [1]{\endgroup\@href {#1}{\urlprefix }}%
\providecommand \urlprefix  [0]{URL }%
\providecommand \Eprint [0]{\href }%
\providecommand \doibase [0]{http://dx.doi.org/}%
\providecommand \selectlanguage [0]{\@gobble}%
\providecommand \bibinfo  [0]{\@secondoftwo}%
\providecommand \bibfield  [0]{\@secondoftwo}%
\providecommand \translation [1]{[#1]}%
\providecommand \BibitemOpen [0]{}%
\providecommand \bibitemStop [0]{}%
\providecommand \bibitemNoStop [0]{.\EOS\space}%
\providecommand \EOS [0]{\spacefactor3000\relax}%
\providecommand \BibitemShut  [1]{\csname bibitem#1\endcsname}%
\let\auto@bib@innerbib\@empty
\bibitem [{\citenamefont {Zurek}(2003)}]{ZurekRMP}%
  \BibitemOpen
  \bibfield  {author} {\bibinfo {author} {\bibfnamefont {W.~H.}\ \bibnamefont
  {Zurek}},\ }\bibfield  {title} {\enquote {\bibinfo {title} {Decoherence,
  einselection, and the quantum origins of the classical},}\ }\href {\doibase
  10.1103/RevModPhys.75.715} {\bibfield  {journal} {\bibinfo  {journal} {Rev.
  Mod. Phys.}\ }\textbf {\bibinfo {volume} {75}},\ \bibinfo {pages} {715--775}
  (\bibinfo {year} {2003})}\BibitemShut {NoStop}%
\bibitem [{\citenamefont {Zurek}(2009)}]{ZurekNatPhys2009}%
  \BibitemOpen
  \bibfield  {author} {\bibinfo {author} {\bibfnamefont {W.~H.}\ \bibnamefont
  {Zurek}},\ }\bibfield  {title} {\enquote {\bibinfo {title} {Quantum
  darwinism},}\ }\href {\doibase 10.1038/nphys1202} {\bibfield  {journal}
  {\bibinfo  {journal} {Nature Physics}\ }\textbf {\bibinfo {volume} {5}},\
  \bibinfo {pages} {181} (\bibinfo {year} {2009})}\BibitemShut {NoStop}%
\bibitem [{\citenamefont {Blume-Kohout}\ and\ \citenamefont
  {Zurek}(2006)}]{KohoutPRA2006}%
  \BibitemOpen
  \bibfield  {author} {\bibinfo {author} {\bibfnamefont {R.}~\bibnamefont
  {Blume-Kohout}}\ and\ \bibinfo {author} {\bibfnamefont {W.~H.}\ \bibnamefont
  {Zurek}},\ }\bibfield  {title} {\enquote {\bibinfo {title} {Quantum
  darwinism: Entanglement, branches, and the emergent classicality of
  redundantly stored quantum information},}\ }\href {\doibase
  10.1103/PhysRevA.73.062310} {\bibfield  {journal} {\bibinfo  {journal} {Phys.
  Rev. A}\ }\textbf {\bibinfo {volume} {73}},\ \bibinfo {pages} {062310}
  (\bibinfo {year} {2006})}\BibitemShut {NoStop}%
\bibitem [{\citenamefont {Le}\ and\ \citenamefont
  {Olaya-Castro}(2019)}]{LePRL2018}%
  \BibitemOpen
  \bibfield  {author} {\bibinfo {author} {\bibfnamefont {T.~P.}\ \bibnamefont
  {Le}}\ and\ \bibinfo {author} {\bibfnamefont {A.}~\bibnamefont
  {Olaya-Castro}},\ }\bibfield  {title} {\enquote {\bibinfo {title} {Strong
  quantum darwinism and strong independence are equivalent to spectrum
  broadcast structure},}\ }\href {\doibase 10.1103/PhysRevLett.122.010403}
  {\bibfield  {journal} {\bibinfo  {journal} {Phys. Rev. Lett.}\ }\textbf
  {\bibinfo {volume} {122}},\ \bibinfo {pages} {010403} (\bibinfo {year}
  {2019})}\BibitemShut {NoStop}%
\bibitem [{\citenamefont {Le}\ and\ \citenamefont
  {Olaya-Castro}(2018)}]{LePRA2018}%
  \BibitemOpen
  \bibfield  {author} {\bibinfo {author} {\bibfnamefont {T.~P.}\ \bibnamefont
  {Le}}\ and\ \bibinfo {author} {\bibfnamefont {A.}~\bibnamefont
  {Olaya-Castro}},\ }\bibfield  {title} {\enquote {\bibinfo {title}
  {Objectivity (or lack thereof): Comparison between predictions of quantum
  {D}arwinism and spectrum broadcast structure},}\ }\href {\doibase
  10.1103/PhysRevA.98.032103} {\bibfield  {journal} {\bibinfo  {journal} {Phys.
  Rev. A}\ }\textbf {\bibinfo {volume} {98}},\ \bibinfo {pages} {032103}
  (\bibinfo {year} {2018})}\BibitemShut {NoStop}%
\bibitem [{\citenamefont {Le}\ and\ \citenamefont
  {Olaya-Castro}(2020)}]{Le2020}%
  \BibitemOpen
  \bibfield  {author} {\bibinfo {author} {\bibfnamefont {T.~P.}\ \bibnamefont
  {Le}}\ and\ \bibinfo {author} {\bibfnamefont {A.}~\bibnamefont
  {Olaya-Castro}},\ }\bibfield  {title} {\enquote {\bibinfo {title} {Witnessing
  non-objectivity in the framework of strong quantum darwinism},}\ }\href
  {\doibase 10.1103/PhysRevLett.122.010403} {\bibfield  {journal} {\bibinfo
  {journal} {Quantum Sci. Technol.}\ }\textbf {\bibinfo {volume} {5}},\
  \bibinfo {pages} {045012} (\bibinfo {year} {2020})}\BibitemShut {NoStop}%
\bibitem [{\citenamefont {Horodecki}\ \emph {et~al.}(2015)\citenamefont
  {Horodecki}, \citenamefont {Korbicz},\ and\ \citenamefont
  {Horodecki}}]{HorodeckiPRA2015}%
  \BibitemOpen
  \bibfield  {author} {\bibinfo {author} {\bibfnamefont {R.}~\bibnamefont
  {Horodecki}}, \bibinfo {author} {\bibfnamefont {J.~K.}\ \bibnamefont
  {Korbicz}}, \ and\ \bibinfo {author} {\bibfnamefont {P.}~\bibnamefont
  {Horodecki}},\ }\bibfield  {title} {\enquote {\bibinfo {title} {Quantum
  origins of objectivity},}\ }\href {\doibase 10.1103/PhysRevA.91.032122}
  {\bibfield  {journal} {\bibinfo  {journal} {Phys. Rev. A}\ }\textbf {\bibinfo
  {volume} {91}},\ \bibinfo {pages} {032122} (\bibinfo {year}
  {2015})}\BibitemShut {NoStop}%
\bibitem [{\citenamefont {Korbicz}\ \emph {et~al.}(2014)\citenamefont
  {Korbicz}, \citenamefont {Horodecki},\ and\ \citenamefont
  {Horodecki}}]{KorbiczPRL2014}%
  \BibitemOpen
  \bibfield  {author} {\bibinfo {author} {\bibfnamefont {J.~K.}\ \bibnamefont
  {Korbicz}}, \bibinfo {author} {\bibfnamefont {P.}~\bibnamefont {Horodecki}},
  \ and\ \bibinfo {author} {\bibfnamefont {R.}~\bibnamefont {Horodecki}},\
  }\bibfield  {title} {\enquote {\bibinfo {title} {Objectivity in a noisy
  photonic environment through quantum state information broadcasting},}\
  }\href {\doibase 10.1103/PhysRevLett.112.120402} {\bibfield  {journal}
  {\bibinfo  {journal} {Phys. Rev. Lett.}\ }\textbf {\bibinfo {volume} {112}},\
  \bibinfo {pages} {120402} (\bibinfo {year} {2014})}\BibitemShut {NoStop}%
\bibitem [{\citenamefont {Korbicz}\ \emph {et~al.}(2017)\citenamefont
  {Korbicz}, \citenamefont {Aguilar}, \citenamefont {\ifmmode \acute{C}\else
  \'{C}\fi{}wikli\ifmmode~\acute{n}\else \'{n}\fi{}ski},\ and\ \citenamefont
  {Horodecki}}]{KorbiczPRA2017}%
  \BibitemOpen
  \bibfield  {author} {\bibinfo {author} {\bibfnamefont {J.~K.}\ \bibnamefont
  {Korbicz}}, \bibinfo {author} {\bibfnamefont {E.~A.}\ \bibnamefont
  {Aguilar}}, \bibinfo {author} {\bibfnamefont {P.}~\bibnamefont {\ifmmode
  \acute{C}\else \'{C}\fi{}wikli\ifmmode~\acute{n}\else \'{n}\fi{}ski}}, \ and\
  \bibinfo {author} {\bibfnamefont {P.}~\bibnamefont {Horodecki}},\ }\bibfield
  {title} {\enquote {\bibinfo {title} {Generic appearance of objective results
  in quantum measurements},}\ }\href {\doibase 10.1103/PhysRevA.96.032124}
  {\bibfield  {journal} {\bibinfo  {journal} {Phys. Rev. A}\ }\textbf {\bibinfo
  {volume} {96}},\ \bibinfo {pages} {032124} (\bibinfo {year}
  {2017})}\BibitemShut {NoStop}%
\bibitem [{\citenamefont {Zwolak}\ and\ \citenamefont
  {Zurek}(2013)}]{ZurekSciRep2013}%
  \BibitemOpen
  \bibfield  {author} {\bibinfo {author} {\bibfnamefont {M.}~\bibnamefont
  {Zwolak}}\ and\ \bibinfo {author} {\bibfnamefont {W.~H.}\ \bibnamefont
  {Zurek}},\ }\bibfield  {title} {\enquote {\bibinfo {title} {Complementarity
  of quantum discord and classically accessible information},}\ }\href
  {\doibase 10.1038/srep01729} {\bibfield  {journal} {\bibinfo  {journal} {Sci.
  Rep.}\ }\textbf {\bibinfo {volume} {3}},\ \bibinfo {pages} {1729} (\bibinfo
  {year} {2013})}\BibitemShut {NoStop}%
\bibitem [{\citenamefont {Zwolak}\ and\ \citenamefont
  {Zurek}(2016)}]{ZurekSciRep2016}%
  \BibitemOpen
  \bibfield  {author} {\bibinfo {author} {\bibfnamefont {M.}~\bibnamefont
  {Zwolak}}\ and\ \bibinfo {author} {\bibfnamefont {W.~H.}\ \bibnamefont
  {Zurek}},\ }\bibfield  {title} {\enquote {\bibinfo {title} {Amplification,
  decoherence, and the acquisition of information by spin environments},}\
  }\href {\doibase 10.1038/srep25277} {\bibfield  {journal} {\bibinfo
  {journal} {Sci. Rep.}\ }\textbf {\bibinfo {volume} {6}},\ \bibinfo {pages}
  {25277} (\bibinfo {year} {2016})}\BibitemShut {NoStop}%
\bibitem [{\citenamefont {Zwolak}\ and\ \citenamefont
  {Zurek}(2017)}]{ZurekPRA2017}%
  \BibitemOpen
  \bibfield  {author} {\bibinfo {author} {\bibfnamefont {M.}~\bibnamefont
  {Zwolak}}\ and\ \bibinfo {author} {\bibfnamefont {W.~H.}\ \bibnamefont
  {Zurek}},\ }\bibfield  {title} {\enquote {\bibinfo {title} {Redundancy of
  einselected information in quantum darwinism: The irrelevance of irrelevant
  environment bits},}\ }\href {\doibase 10.1103/PhysRevA.95.030101} {\bibfield
  {journal} {\bibinfo  {journal} {Phys. Rev. A}\ }\textbf {\bibinfo {volume}
  {95}},\ \bibinfo {pages} {030101(R)} (\bibinfo {year} {2017})}\BibitemShut
  {NoStop}%
\bibitem [{\citenamefont {Knott}\ \emph {et~al.}(2018)\citenamefont {Knott},
  \citenamefont {Tufarelli}, \citenamefont {Piani},\ and\ \citenamefont
  {Adesso}}]{AdessoPRL2018}%
  \BibitemOpen
  \bibfield  {author} {\bibinfo {author} {\bibfnamefont {P.~A.}\ \bibnamefont
  {Knott}}, \bibinfo {author} {\bibfnamefont {T.}~\bibnamefont {Tufarelli}},
  \bibinfo {author} {\bibfnamefont {M.}~\bibnamefont {Piani}}, \ and\ \bibinfo
  {author} {\bibfnamefont {G.}~\bibnamefont {Adesso}},\ }\bibfield  {title}
  {\enquote {\bibinfo {title} {Generic emergence of objectivity of observables
  in infinite dimensions},}\ }\href {\doibase 10.1103/PhysRevLett.121.160401}
  {\bibfield  {journal} {\bibinfo  {journal} {Phys. Rev. Lett.}\ }\textbf
  {\bibinfo {volume} {121}},\ \bibinfo {pages} {160401} (\bibinfo {year}
  {2018})}\BibitemShut {NoStop}%
\bibitem [{\citenamefont {Balaneskovica}(2015)}]{BalaneskovicaEPJD2015}%
  \BibitemOpen
  \bibfield  {author} {\bibinfo {author} {\bibfnamefont {N.}~\bibnamefont
  {Balaneskovica}},\ }\bibfield  {title} {\enquote {\bibinfo {title} {Random
  unitary evolution model of quantum darwinism with pure decoherence},}\ }\href
  {\doibase 10.1140/epjd/e2015-60319-9} {\bibfield  {journal} {\bibinfo
  {journal} {Eur. Phys. J. D}\ }\textbf {\bibinfo {volume} {69}},\ \bibinfo
  {pages} {232} (\bibinfo {year} {2015})}\BibitemShut {NoStop}%
\bibitem [{\citenamefont {Balaneskovica}\ and\ \citenamefont
  {Mendler}(2016)}]{MendlerEPJD2016}%
  \BibitemOpen
  \bibfield  {author} {\bibinfo {author} {\bibfnamefont {N.}~\bibnamefont
  {Balaneskovica}}\ and\ \bibinfo {author} {\bibfnamefont {M.}~\bibnamefont
  {Mendler}},\ }\bibfield  {title} {\enquote {\bibinfo {title} {Dissipation,
  dephasing and quantum darwinism in qubit systems with random unitary
  interactions},}\ }\href {\doibase 10.1140/epjd/e2016-70174-9} {\bibfield
  {journal} {\bibinfo  {journal} {Eur. Phys. J. D}\ }\textbf {\bibinfo {volume}
  {70}},\ \bibinfo {pages} {177} (\bibinfo {year} {2016})}\BibitemShut
  {NoStop}%
\bibitem [{\citenamefont {Lorenzo}\ \emph {et~al.}(2020)\citenamefont
  {Lorenzo}, \citenamefont {Paternostro},\ and\ \citenamefont
  {Palma}}]{SalvatorePRR}%
  \BibitemOpen
  \bibfield  {author} {\bibinfo {author} {\bibfnamefont {S.}~\bibnamefont
  {Lorenzo}}, \bibinfo {author} {\bibfnamefont {M.}~\bibnamefont
  {Paternostro}}, \ and\ \bibinfo {author} {\bibfnamefont {G.~M.}\ \bibnamefont
  {Palma}},\ }\bibfield  {title} {\enquote {\bibinfo {title} {Anti-zeno-based
  dynamical control of the unfolding of quantum darwinism},}\ }\href {\doibase
  10.1103/PhysRevResearch.2.013164} {\bibfield  {journal} {\bibinfo  {journal}
  {Phys. Rev. Research}\ }\textbf {\bibinfo {volume} {2}},\ \bibinfo {pages}
  {013164} (\bibinfo {year} {2020})}\BibitemShut {NoStop}%
\bibitem [{\citenamefont {Garc\'{\i}a-P\'erez}\ \emph
  {et~al.}(2020{\natexlab{a}})\citenamefont {Garc\'{\i}a-P\'erez},
  \citenamefont {Chisholm}, \citenamefont {Rossi}, \citenamefont {Palma},\ and\
  \citenamefont {Maniscalco}}]{SabrinaPRR}%
  \BibitemOpen
  \bibfield  {author} {\bibinfo {author} {\bibfnamefont {G.}~\bibnamefont
  {Garc\'{\i}a-P\'erez}}, \bibinfo {author} {\bibfnamefont {D.~A.}\
  \bibnamefont {Chisholm}}, \bibinfo {author} {\bibfnamefont {M.~A.~C.}\
  \bibnamefont {Rossi}}, \bibinfo {author} {\bibfnamefont {G.~M.}\ \bibnamefont
  {Palma}}, \ and\ \bibinfo {author} {\bibfnamefont {S.}~\bibnamefont
  {Maniscalco}},\ }\bibfield  {title} {\enquote {\bibinfo {title} {Decoherence
  without entanglement and quantum {D}arwinism},}\ }\href {\doibase
  10.1103/PhysRevResearch.2.012061} {\bibfield  {journal} {\bibinfo  {journal}
  {Phys. Rev. Research}\ }\textbf {\bibinfo {volume} {2}},\ \bibinfo {pages}
  {012061} (\bibinfo {year} {2020}{\natexlab{a}})}\BibitemShut {NoStop}%
\bibitem [{\citenamefont {Pleasance}\ and\ \citenamefont
  {Garraway}(2017)}]{GarrawayPRA2017}%
  \BibitemOpen
  \bibfield  {author} {\bibinfo {author} {\bibfnamefont {G.}~\bibnamefont
  {Pleasance}}\ and\ \bibinfo {author} {\bibfnamefont {B.~M.}\ \bibnamefont
  {Garraway}},\ }\bibfield  {title} {\enquote {\bibinfo {title} {Application of
  quantum darwinism to a structured environment},}\ }\href {\doibase
  10.1103/PhysRevA.96.062105} {\bibfield  {journal} {\bibinfo  {journal} {Phys.
  Rev. A}\ }\textbf {\bibinfo {volume} {96}},\ \bibinfo {pages} {062105}
  (\bibinfo {year} {2017})}\BibitemShut {NoStop}%
\bibitem [{\citenamefont {Zwolak}\ \emph {et~al.}(2010)\citenamefont {Zwolak},
  \citenamefont {Quan},\ and\ \citenamefont {Zurek}}]{ZwolakPRA2010}%
  \BibitemOpen
  \bibfield  {author} {\bibinfo {author} {\bibfnamefont {M.}~\bibnamefont
  {Zwolak}}, \bibinfo {author} {\bibfnamefont {H.~T.}\ \bibnamefont {Quan}}, \
  and\ \bibinfo {author} {\bibfnamefont {W.~H.}\ \bibnamefont {Zurek}},\
  }\bibfield  {title} {\enquote {\bibinfo {title} {Redundant imprinting of
  information in nonideal environments: Objective reality via a noisy
  channel},}\ }\href {\doibase 10.1103/PhysRevA.81.062110} {\bibfield
  {journal} {\bibinfo  {journal} {Phys. Rev. A}\ }\textbf {\bibinfo {volume}
  {81}},\ \bibinfo {pages} {062110} (\bibinfo {year} {2010})}\BibitemShut
  {NoStop}%
\bibitem [{\citenamefont {Zwolak}\ \emph {et~al.}(2009)\citenamefont {Zwolak},
  \citenamefont {Quan},\ and\ \citenamefont {Zurek}}]{ZwolakPRL2009}%
  \BibitemOpen
  \bibfield  {author} {\bibinfo {author} {\bibfnamefont {M.}~\bibnamefont
  {Zwolak}}, \bibinfo {author} {\bibfnamefont {H.~T.}\ \bibnamefont {Quan}}, \
  and\ \bibinfo {author} {\bibfnamefont {W.~H.}\ \bibnamefont {Zurek}},\
  }\bibfield  {title} {\enquote {\bibinfo {title} {Quantum darwinism in a mixed
  environment},}\ }\href {\doibase 10.1103/PhysRevLett.103.110402} {\bibfield
  {journal} {\bibinfo  {journal} {Phys. Rev. Lett.}\ }\textbf {\bibinfo
  {volume} {103}},\ \bibinfo {pages} {110402} (\bibinfo {year}
  {2009})}\BibitemShut {NoStop}%
\bibitem [{\citenamefont {Touil}\ \emph {et~al.}(2021)\citenamefont {Touil},
  \citenamefont {Yan}, \citenamefont {Girolami}, \citenamefont {Deffner},\ and\
  \citenamefont {Zurek}}]{AkramPRL}%
  \BibitemOpen
  \bibfield  {author} {\bibinfo {author} {\bibfnamefont {A.}~\bibnamefont
  {Touil}}, \bibinfo {author} {\bibfnamefont {B.}~\bibnamefont {Yan}}, \bibinfo
  {author} {\bibfnamefont {D.}~\bibnamefont {Girolami}}, \bibinfo {author}
  {\bibfnamefont {S.}~\bibnamefont {Deffner}}, \ and\ \bibinfo {author}
  {\bibfnamefont {W.~H.}\ \bibnamefont {Zurek}},\ }\bibfield  {title} {\enquote
  {\bibinfo {title} {Eavesdropping on the decohering environment: Quantum
  darwinism, amplification, and the origin of objective classical reality},}\
  }\href {\doibase 10.1103/PhysRevLett.128.010401} {\bibfield  {journal}
  {\bibinfo  {journal} {Phys. Rev. Lett.}\ }\textbf {\bibinfo {volume} {128}},\
  \bibinfo {pages} {010401} (\bibinfo {year} {2021})}\BibitemShut {NoStop}%
\bibitem [{\citenamefont {Mirkin}\ and\ \citenamefont
  {Wisniacki}(2021)}]{MirkinEntropy}%
  \BibitemOpen
  \bibfield  {author} {\bibinfo {author} {\bibfnamefont {N.}~\bibnamefont
  {Mirkin}}\ and\ \bibinfo {author} {\bibfnamefont {D.~A.}\ \bibnamefont
  {Wisniacki}},\ }\bibfield  {title} {\enquote {\bibinfo {title} {Many-body
  localization and the emergence of quantum darwinism},}\ }\href {\doibase
  10.3390/e23111506} {\bibfield  {journal} {\bibinfo  {journal} {Entropy}\
  }\textbf {\bibinfo {volume} {23}},\ \bibinfo {pages} {1377} (\bibinfo {year}
  {2021})}\BibitemShut {NoStop}%
\bibitem [{\citenamefont {\ifmmode~\mbox{\c{C}}\else \c{C}\fi{}akmak}\ \emph
  {et~al.}(2021)\citenamefont {\ifmmode~\mbox{\c{C}}\else \c{C}\fi{}akmak},
  \citenamefont {M\"ustecapl\ifmmode \imath \else \i
  \fi{}o\ifmmode~\breve{g}\else \u{g}\fi{}lu}, \citenamefont {Paternostro},
  \citenamefont {Vacchini},\ and\ \citenamefont {Campbell}}]{CakmakEntropy}%
  \BibitemOpen
  \bibfield  {author} {\bibinfo {author} {\bibfnamefont {B.}~\bibnamefont
  {\ifmmode~\mbox{\c{C}}\else \c{C}\fi{}akmak}}, \bibinfo {author}
  {\bibfnamefont {\"O.~E.}\ \bibnamefont {M\"ustecapl\ifmmode \imath \else \i
  \fi{}o\ifmmode~\breve{g}\else \u{g}\fi{}lu}}, \bibinfo {author}
  {\bibfnamefont {M.}~\bibnamefont {Paternostro}}, \bibinfo {author}
  {\bibfnamefont {B.}~\bibnamefont {Vacchini}}, \ and\ \bibinfo {author}
  {\bibfnamefont {S.}~\bibnamefont {Campbell}},\ }\bibfield  {title} {\enquote
  {\bibinfo {title} {Quantum darwinism in a composite system: Objectivity
  versus classicality},}\ }\href {\doibase 10.3390/e23080995} {\bibfield
  {journal} {\bibinfo  {journal} {Entropy}\ }\textbf {\bibinfo {volume} {23}},\
  \bibinfo {pages} {995} (\bibinfo {year} {2021})}\BibitemShut {NoStop}%
\bibitem [{\citenamefont {Ryan}\ \emph {et~al.}(2021)\citenamefont {Ryan},
  \citenamefont {Paternostro},\ and\ \citenamefont {Campbell}}]{RyanPLA}%
  \BibitemOpen
  \bibfield  {author} {\bibinfo {author} {\bibfnamefont {E.}~\bibnamefont
  {Ryan}}, \bibinfo {author} {\bibfnamefont {M.}~\bibnamefont {Paternostro}}, \
  and\ \bibinfo {author} {\bibfnamefont {S.}~\bibnamefont {Campbell}},\
  }\bibfield  {title} {\enquote {\bibinfo {title} {Quantum darwinism in a
  structured spin environment},}\ }\href {\doibase
  10.1016/j.physleta.2021.127675} {\bibfield  {journal} {\bibinfo  {journal}
  {Phys. Lett. A}\ }\textbf {\bibinfo {volume} {416}},\ \bibinfo {pages}
  {127675} (\bibinfo {year} {2021})}\BibitemShut {NoStop}%
\bibitem [{\citenamefont {Riedel}\ \emph {et~al.}(2012)\citenamefont {Riedel},
  \citenamefont {Zurek},\ and\ \citenamefont {Zwolak}}]{ZwolakNJP2012}%
  \BibitemOpen
  \bibfield  {author} {\bibinfo {author} {\bibfnamefont {C.~Jess}\ \bibnamefont
  {Riedel}}, \bibinfo {author} {\bibfnamefont {W.~H.}\ \bibnamefont {Zurek}}, \
  and\ \bibinfo {author} {\bibfnamefont {M.}~\bibnamefont {Zwolak}},\
  }\bibfield  {title} {\enquote {\bibinfo {title} {The rise and fall of
  redundancy in decoherence and quantum darwinism},}\ }\href {\doibase
  10.1088/1367-2630/14/8/083010} {\bibfield  {journal} {\bibinfo  {journal}
  {New J. Phys.}\ }\textbf {\bibinfo {volume} {14}},\ \bibinfo {pages} {083010}
  (\bibinfo {year} {2012})}\BibitemShut {NoStop}%
\bibitem [{\citenamefont {Campbell}\ \emph {et~al.}(2019)\citenamefont
  {Campbell}, \citenamefont {\ifmmode~\mbox{\c{C}}\else \c{C}\fi{}akmak},
  \citenamefont {M\"ustecapl\ifmmode \imath \else \i
  \fi{}o\ifmmode~\breve{g}\else \u{g}\fi{}lu}, \citenamefont {Paternostro},\
  and\ \citenamefont {Vacchini}}]{CampbellPRA2019}%
  \BibitemOpen
  \bibfield  {author} {\bibinfo {author} {\bibfnamefont {S.}~\bibnamefont
  {Campbell}}, \bibinfo {author} {\bibfnamefont {B.}~\bibnamefont
  {\ifmmode~\mbox{\c{C}}\else \c{C}\fi{}akmak}}, \bibinfo {author}
  {\bibfnamefont {\"O.~E.}\ \bibnamefont {M\"ustecapl\ifmmode \imath \else \i
  \fi{}o\ifmmode~\breve{g}\else \u{g}\fi{}lu}}, \bibinfo {author}
  {\bibfnamefont {M.}~\bibnamefont {Paternostro}}, \ and\ \bibinfo {author}
  {\bibfnamefont {B.}~\bibnamefont {Vacchini}},\ }\bibfield  {title} {\enquote
  {\bibinfo {title} {Collisional unfolding of quantum darwinism},}\ }\href
  {\doibase 10.1103/PhysRevA.99.042103} {\bibfield  {journal} {\bibinfo
  {journal} {Phys. Rev. A}\ }\textbf {\bibinfo {volume} {99}},\ \bibinfo
  {pages} {042103} (\bibinfo {year} {2019})}\BibitemShut {NoStop}%
\bibitem [{\citenamefont {Le}\ \emph {et~al.}(2021)\citenamefont {Le},
  \citenamefont {Adesso},\ and\ \citenamefont {Winter}}]{LeEntropy}%
  \BibitemOpen
  \bibfield  {author} {\bibinfo {author} {\bibfnamefont {T.~P.}\ \bibnamefont
  {Le}}, \bibinfo {author} {\bibfnamefont {G.}~\bibnamefont {Adesso}}, \ and\
  \bibinfo {author} {\bibfnamefont {A.}~\bibnamefont {Winter}},\ }\bibfield
  {title} {\enquote {\bibinfo {title} {Thermality versus objectivity: Can they
  peacefully coexist?}}\ }\href {\doibase 10.3390/e23111506} {\bibfield
  {journal} {\bibinfo  {journal} {Entropy}\ }\textbf {\bibinfo {volume} {23}},\
  \bibinfo {pages} {1506} (\bibinfo {year} {2021})}\BibitemShut {NoStop}%
\bibitem [{\citenamefont {Roszak}\ and\ \citenamefont
  {Korbicz}(2019)}]{KorbiczPRA19}%
  \BibitemOpen
  \bibfield  {author} {\bibinfo {author} {\bibfnamefont {K.}~\bibnamefont
  {Roszak}}\ and\ \bibinfo {author} {\bibfnamefont {J.~K.}\ \bibnamefont
  {Korbicz}},\ }\bibfield  {title} {\enquote {\bibinfo {title} {Entanglement
  and objectivity in pure dephasing models},}\ }\href {\doibase
  10.1103/PhysRevA.100.062127} {\bibfield  {journal} {\bibinfo  {journal}
  {Phys. Rev. A}\ }\textbf {\bibinfo {volume} {100}},\ \bibinfo {pages}
  {062127} (\bibinfo {year} {2019})}\BibitemShut {NoStop}%
\bibitem [{\citenamefont {Korbicz}(2021)}]{Korbicz_Quantum}%
  \BibitemOpen
  \bibfield  {author} {\bibinfo {author} {\bibfnamefont {J.~K.}\ \bibnamefont
  {Korbicz}},\ }\bibfield  {title} {\enquote {\bibinfo {title} {Roads to
  objectivity: Quantum {D}arwinism, {S}pectrum {B}roadcast {S}tructures, and
  {S}trong quantum {D}arwinism},}\ }\href {\doibase 10.22331/q-2021-11-08-571}
  {\bibfield  {journal} {\bibinfo  {journal} {Quantum}\ }\textbf {\bibinfo
  {volume} {5}},\ \bibinfo {pages} {571} (\bibinfo {year} {2021})}\BibitemShut
  {NoStop}%
\bibitem [{\citenamefont {Tuziemski}\ \emph {et~al.}(2019)\citenamefont
  {Tuziemski}, \citenamefont {Lampo}, \citenamefont {Lewenstein},\ and\
  \citenamefont {Korbicz}}]{KorbiczPRA2019b}%
  \BibitemOpen
  \bibfield  {author} {\bibinfo {author} {\bibfnamefont {J.}~\bibnamefont
  {Tuziemski}}, \bibinfo {author} {\bibfnamefont {A.}~\bibnamefont {Lampo}},
  \bibinfo {author} {\bibfnamefont {M.}~\bibnamefont {Lewenstein}}, \ and\
  \bibinfo {author} {\bibfnamefont {J.~K.}\ \bibnamefont {Korbicz}},\
  }\bibfield  {title} {\enquote {\bibinfo {title} {Reexamination of the
  decoherence of spin registers},}\ }\href {\doibase
  10.1103/PhysRevA.99.022122} {\bibfield  {journal} {\bibinfo  {journal} {Phys.
  Rev. A}\ }\textbf {\bibinfo {volume} {99}},\ \bibinfo {pages} {022122}
  (\bibinfo {year} {2019})}\BibitemShut {NoStop}%
\bibitem [{\citenamefont {Kici\ifmmode~\acute{n}\else \'{n}\fi{}ski}\ and\
  \citenamefont {Korbicz}(2021)}]{KorbiczPRA2021}%
  \BibitemOpen
  \bibfield  {author} {\bibinfo {author} {\bibfnamefont {M.}~\bibnamefont
  {Kici\ifmmode~\acute{n}\else \'{n}\fi{}ski}}\ and\ \bibinfo {author}
  {\bibfnamefont {J.~K.}\ \bibnamefont {Korbicz}},\ }\bibfield  {title}
  {\enquote {\bibinfo {title} {Decoherence and objectivity in higher spin
  environments},}\ }\href {\doibase 10.1103/PhysRevA.104.042216} {\bibfield
  {journal} {\bibinfo  {journal} {Phys. Rev. A}\ }\textbf {\bibinfo {volume}
  {104}},\ \bibinfo {pages} {042216} (\bibinfo {year} {2021})}\BibitemShut
  {NoStop}%
\bibitem [{\citenamefont {Ciampini}\ \emph {et~al.}(2018)\citenamefont
  {Ciampini}, \citenamefont {Pinna}, \citenamefont {Mataloni},\ and\
  \citenamefont {Paternostro}}]{DarwinismExp1}%
  \BibitemOpen
  \bibfield  {author} {\bibinfo {author} {\bibfnamefont {M.~A.}\ \bibnamefont
  {Ciampini}}, \bibinfo {author} {\bibfnamefont {G.}~\bibnamefont {Pinna}},
  \bibinfo {author} {\bibfnamefont {P.}~\bibnamefont {Mataloni}}, \ and\
  \bibinfo {author} {\bibfnamefont {M.}~\bibnamefont {Paternostro}},\
  }\bibfield  {title} {\enquote {\bibinfo {title} {Experimental signature of
  quantum darwinism in photonic cluster states},}\ }\href {\doibase
  10.1103/PhysRevA.98.020101} {\bibfield  {journal} {\bibinfo  {journal} {Phys.
  Rev. A}\ }\textbf {\bibinfo {volume} {98}},\ \bibinfo {pages} {020101(R)}
  (\bibinfo {year} {2018})}\BibitemShut {NoStop}%
\bibitem [{\citenamefont {Chen}\ \emph {et~al.}(2019)\citenamefont {Chen},
  \citenamefont {Zhong}, \citenamefont {Li}, \citenamefont {Wu}, \citenamefont
  {Wang}, \citenamefont {Li}, \citenamefont {Liu}, \citenamefont {Lu},\ and\
  \citenamefont {Pan}}]{DarwinismExp2}%
  \BibitemOpen
  \bibfield  {author} {\bibinfo {author} {\bibfnamefont {M.-C.}\ \bibnamefont
  {Chen}}, \bibinfo {author} {\bibfnamefont {H.-S.}\ \bibnamefont {Zhong}},
  \bibinfo {author} {\bibfnamefont {Y.}~\bibnamefont {Li}}, \bibinfo {author}
  {\bibfnamefont {D.}~\bibnamefont {Wu}}, \bibinfo {author} {\bibfnamefont
  {X.-L.}\ \bibnamefont {Wang}}, \bibinfo {author} {\bibfnamefont
  {L.}~\bibnamefont {Li}}, \bibinfo {author} {\bibfnamefont {N.-L.}\
  \bibnamefont {Liu}}, \bibinfo {author} {\bibfnamefont {C.-Y.}\ \bibnamefont
  {Lu}}, \ and\ \bibinfo {author} {\bibfnamefont {J.-W.}\ \bibnamefont {Pan}},\
  }\bibfield  {title} {\enquote {\bibinfo {title} {Emergence of classical
  objectivity on a quantum darwinism simulator},}\ }\href {\doibase
  10.1016/j.scib.2019.03.032} {\bibfield  {journal} {\bibinfo  {journal}
  {Science Bulletin}\ }\textbf {\bibinfo {volume} {64}},\ \bibinfo {pages}
  {580} (\bibinfo {year} {2019})}\BibitemShut {NoStop}%
\bibitem [{\citenamefont {Unden}\ \emph {et~al.}(2019)\citenamefont {Unden},
  \citenamefont {Louzon}, \citenamefont {Zwolak}, \citenamefont {Zurek},\ and\
  \citenamefont {Jelezko}}]{DarwinismExp3}%
  \BibitemOpen
  \bibfield  {author} {\bibinfo {author} {\bibfnamefont {T.~K.}\ \bibnamefont
  {Unden}}, \bibinfo {author} {\bibfnamefont {D.}~\bibnamefont {Louzon}},
  \bibinfo {author} {\bibfnamefont {M.}~\bibnamefont {Zwolak}}, \bibinfo
  {author} {\bibfnamefont {W.~H.}\ \bibnamefont {Zurek}}, \ and\ \bibinfo
  {author} {\bibfnamefont {F.}~\bibnamefont {Jelezko}},\ }\bibfield  {title}
  {\enquote {\bibinfo {title} {Revealing the emergence of classicality using
  nitrogen-vacancy centers},}\ }\href {\doibase 10.1103/PhysRevLett.123.140402}
  {\bibfield  {journal} {\bibinfo  {journal} {Phys. Rev. Lett.}\ }\textbf
  {\bibinfo {volume} {123}},\ \bibinfo {pages} {140402} (\bibinfo {year}
  {2019})}\BibitemShut {NoStop}%
\bibitem [{\citenamefont {Lampo}\ \emph {et~al.}(2017)\citenamefont {Lampo},
  \citenamefont {Tuziemski}, \citenamefont {Lewenstein},\ and\ \citenamefont
  {Korbicz}}]{LewensteinPRA2017}%
  \BibitemOpen
  \bibfield  {author} {\bibinfo {author} {\bibfnamefont {A.}~\bibnamefont
  {Lampo}}, \bibinfo {author} {\bibfnamefont {J.}~\bibnamefont {Tuziemski}},
  \bibinfo {author} {\bibfnamefont {M.}~\bibnamefont {Lewenstein}}, \ and\
  \bibinfo {author} {\bibfnamefont {J.~K.}\ \bibnamefont {Korbicz}},\
  }\bibfield  {title} {\enquote {\bibinfo {title} {Objectivity in the
  non-markovian spin-boson model},}\ }\href {\doibase
  10.1103/PhysRevA.96.012120} {\bibfield  {journal} {\bibinfo  {journal} {Phys.
  Rev. A}\ }\textbf {\bibinfo {volume} {96}},\ \bibinfo {pages} {012120}
  (\bibinfo {year} {2017})}\BibitemShut {NoStop}%
\bibitem [{\citenamefont {Galve}\ \emph {et~al.}(2016)\citenamefont {Galve},
  \citenamefont {Zambrini},\ and\ \citenamefont
  {Maniscalco}}]{GalveSciRep2016}%
  \BibitemOpen
  \bibfield  {author} {\bibinfo {author} {\bibfnamefont {F.}~\bibnamefont
  {Galve}}, \bibinfo {author} {\bibfnamefont {R.}~\bibnamefont {Zambrini}}, \
  and\ \bibinfo {author} {\bibfnamefont {S.}~\bibnamefont {Maniscalco}},\
  }\bibfield  {title} {\enquote {\bibinfo {title} {Non-{M}arkovianity hinders
  {Q}uantum {D}arwinism},}\ }\href {\doibase 10.1038/srep19607} {\bibfield
  {journal} {\bibinfo  {journal} {Sci. Rep.}\ }\textbf {\bibinfo {volume}
  {6}},\ \bibinfo {pages} {19607} (\bibinfo {year} {2016})}\BibitemShut
  {NoStop}%
\bibitem [{\citenamefont {Giorgi}\ \emph {et~al.}(2015)\citenamefont {Giorgi},
  \citenamefont {Galve},\ and\ \citenamefont {Zambrini}}]{GiorgiPRA2015}%
  \BibitemOpen
  \bibfield  {author} {\bibinfo {author} {\bibfnamefont {G.~L.}\ \bibnamefont
  {Giorgi}}, \bibinfo {author} {\bibfnamefont {F.}~\bibnamefont {Galve}}, \
  and\ \bibinfo {author} {\bibfnamefont {R.}~\bibnamefont {Zambrini}},\
  }\bibfield  {title} {\enquote {\bibinfo {title} {Quantum {D}arwinism and
  non-{M}arkovian dissipative dynamics from quantum phases of the spin-1/2
  {$XX$} model},}\ }\href {\doibase 10.1103/PhysRevA.92.022105} {\bibfield
  {journal} {\bibinfo  {journal} {Phys. Rev. A}\ }\textbf {\bibinfo {volume}
  {92}},\ \bibinfo {pages} {022105} (\bibinfo {year} {2015})}\BibitemShut
  {NoStop}%
\bibitem [{\citenamefont {Milazzo}\ \emph {et~al.}(2019)\citenamefont
  {Milazzo}, \citenamefont {Lorenzo}, \citenamefont {Paternostro},\ and\
  \citenamefont {Palma}}]{NadiaPRA}%
  \BibitemOpen
  \bibfield  {author} {\bibinfo {author} {\bibfnamefont {N.}~\bibnamefont
  {Milazzo}}, \bibinfo {author} {\bibfnamefont {S.}~\bibnamefont {Lorenzo}},
  \bibinfo {author} {\bibfnamefont {M.}~\bibnamefont {Paternostro}}, \ and\
  \bibinfo {author} {\bibfnamefont {G.~M.}\ \bibnamefont {Palma}},\ }\bibfield
  {title} {\enquote {\bibinfo {title} {Role of information backflow in the
  emergence of quantum darwinism},}\ }\href {\doibase
  10.1103/PhysRevA.100.012101} {\bibfield  {journal} {\bibinfo  {journal}
  {Phys. Rev. A}\ }\textbf {\bibinfo {volume} {100}},\ \bibinfo {pages}
  {012101} (\bibinfo {year} {2019})}\BibitemShut {NoStop}%
\bibitem [{\citenamefont {Megier}\ \emph {et~al.}(2022)\citenamefont {Megier},
  \citenamefont {Smirne}, \citenamefont {Campbell},\ and\ \citenamefont
  {Vacchini}}]{MegierEntropy}%
  \BibitemOpen
  \bibfield  {author} {\bibinfo {author} {\bibfnamefont {N}~\bibnamefont
  {Megier}}, \bibinfo {author} {\bibfnamefont {A}~\bibnamefont {Smirne}},
  \bibinfo {author} {\bibfnamefont {S.}~\bibnamefont {Campbell}}, \ and\
  \bibinfo {author} {\bibfnamefont {B.}~\bibnamefont {Vacchini}},\ }\bibfield
  {title} {\enquote {\bibinfo {title} {Correlations, information backflow, and
  objectivity in a class of pure dephasing models},}\ }\href {\doibase
  10.3390/e24020304} {\bibfield  {journal} {\bibinfo  {journal} {Entropy}\
  }\textbf {\bibinfo {volume} {24}},\ \bibinfo {pages} {304} (\bibinfo {year}
  {2022})}\BibitemShut {NoStop}%
\bibitem [{\citenamefont {Ciccarello}\ \emph {et~al.}(2022)\citenamefont
  {Ciccarello}, \citenamefont {Lorenzo}, \citenamefont {Giovannetti},\ and\
  \citenamefont {Palma}}]{CiccarelloReview}%
  \BibitemOpen
  \bibfield  {author} {\bibinfo {author} {\bibfnamefont {F.}~\bibnamefont
  {Ciccarello}}, \bibinfo {author} {\bibfnamefont {S.}~\bibnamefont {Lorenzo}},
  \bibinfo {author} {\bibfnamefont {V.}~\bibnamefont {Giovannetti}}, \ and\
  \bibinfo {author} {\bibfnamefont {G~M.}\ \bibnamefont {Palma}},\ }\bibfield
  {title} {\enquote {\bibinfo {title} {Quantum collision models: open system
  dynamics from repeated interactions},}\ }\href
  {https://www.sciencedirect.com/science/article/pii/S0370157322000035?via%3Dihub}
  {\bibfield  {journal} {\bibinfo  {journal} {Phys. Rep.}\ }\textbf {\bibinfo
  {volume} {954}},\ \bibinfo {pages} {1} (\bibinfo {year} {2022})}\BibitemShut
  {NoStop}%
\bibitem [{\citenamefont {Campbell}\ and\ \citenamefont
  {Vacchini}(2021)}]{SteveEPL}%
  \BibitemOpen
  \bibfield  {author} {\bibinfo {author} {\bibfnamefont {S.}~\bibnamefont
  {Campbell}}\ and\ \bibinfo {author} {\bibfnamefont {B.}~\bibnamefont
  {Vacchini}},\ }\bibfield  {title} {\enquote {\bibinfo {title} {{Collision
  models in open system dynamics: A versatile tool for deeper insights?}}}\
  }\href {\doibase 10.1209/0295-5075/133/60001} {\bibfield  {journal} {\bibinfo
   {journal} {EPL}\ }\textbf {\bibinfo {volume} {133}},\ \bibinfo {pages}
  {60001} (\bibinfo {year} {2021})}\BibitemShut {NoStop}%
\bibitem [{\citenamefont {Garc\'{\i}a-P\'erez}\ \emph
  {et~al.}(2020{\natexlab{b}})\citenamefont {Garc\'{\i}a-P\'erez},
  \citenamefont {Rossi},\ and\ \citenamefont {Maniscalco}}]{SabrinaNPJQI}%
  \BibitemOpen
  \bibfield  {author} {\bibinfo {author} {\bibfnamefont {G.}~\bibnamefont
  {Garc\'{\i}a-P\'erez}}, \bibinfo {author} {\bibfnamefont {M.~A.~C.}\
  \bibnamefont {Rossi}}, \ and\ \bibinfo {author} {\bibfnamefont
  {S.}~\bibnamefont {Maniscalco}},\ }\bibfield  {title} {\enquote {\bibinfo
  {title} {{IBM} {Q} {E}xperience as a versatile experimental testbed for
  simulating open quantum systems},}\ }\href {\doibase
  10.1038/s41534-019-0235-y} {\bibfield  {journal} {\bibinfo  {journal} {npj
  Quantum Inf.}\ }\textbf {\bibinfo {volume} {6}},\ \bibinfo {pages} {1}
  (\bibinfo {year} {2020}{\natexlab{b}})}\BibitemShut {NoStop}%
\bibitem [{\citenamefont {Lorenzo}\ \emph {et~al.}(2017)\citenamefont
  {Lorenzo}, \citenamefont {Ciccarello},\ and\ \citenamefont
  {Palma}}]{Lorenzo2017}%
  \BibitemOpen
  \bibfield  {author} {\bibinfo {author} {\bibfnamefont {S.}~\bibnamefont
  {Lorenzo}}, \bibinfo {author} {\bibfnamefont {F.}~\bibnamefont {Ciccarello}},
  \ and\ \bibinfo {author} {\bibfnamefont {G.~M.}\ \bibnamefont {Palma}},\
  }\bibfield  {title} {\enquote {\bibinfo {title} {{Composite quantum collision
  models}},}\ }\href {\doibase 10.1103/PhysRevA.96.032107} {\bibfield
  {journal} {\bibinfo  {journal} {Phys. Rev. A}\ }\textbf {\bibinfo {volume}
  {96}},\ \bibinfo {pages} {032107} (\bibinfo {year} {2017})}\BibitemShut
  {NoStop}%
\bibitem [{\citenamefont {Guarnieri}\ \emph {et~al.}(2020)\citenamefont
  {Guarnieri}, \citenamefont {Morrone}, \citenamefont {\c{C}akmak},
  \citenamefont {Plastina},\ and\ \citenamefont {Campbell}}]{GiacomoPLA}%
  \BibitemOpen
  \bibfield  {author} {\bibinfo {author} {\bibfnamefont {G.}~\bibnamefont
  {Guarnieri}}, \bibinfo {author} {\bibfnamefont {D.}~\bibnamefont {Morrone}},
  \bibinfo {author} {\bibfnamefont {B.}~\bibnamefont {\c{C}akmak}}, \bibinfo
  {author} {\bibfnamefont {F.}~\bibnamefont {Plastina}}, \ and\ \bibinfo
  {author} {\bibfnamefont {S.}~\bibnamefont {Campbell}},\ }\bibfield  {title}
  {\enquote {\bibinfo {title} {Non-equilibrium steady-states of memoryless
  quantum collision models},}\ }\href {\doibase 10.1016/j.physleta.2020.126576}
  {\bibfield  {journal} {\bibinfo  {journal} {Phys. Lett. A}\ }\textbf
  {\bibinfo {volume} {384}},\ \bibinfo {pages} {126576} (\bibinfo {year}
  {2020})}\BibitemShut {NoStop}%
\bibitem [{\citenamefont {Baumgratz}\ \emph {et~al.}(2014)\citenamefont
  {Baumgratz}, \citenamefont {Cramer},\ and\ \citenamefont
  {Plenio}}]{Baumgratz}%
  \BibitemOpen
  \bibfield  {author} {\bibinfo {author} {\bibfnamefont {T.}~\bibnamefont
  {Baumgratz}}, \bibinfo {author} {\bibfnamefont {M.}~\bibnamefont {Cramer}}, \
  and\ \bibinfo {author} {\bibfnamefont {M.~B.}\ \bibnamefont {Plenio}},\
  }\bibfield  {title} {\enquote {\bibinfo {title} {Quantifying coherence},}\
  }\href {https://link.aps.org/doi/10.1103/PhysRevLett.113.140401} {\bibfield
  {journal} {\bibinfo  {journal} {Phys. Rev. Lett.}\ }\textbf {\bibinfo
  {volume} {113}},\ \bibinfo {pages} {140401} (\bibinfo {year}
  {2014})}\BibitemShut {NoStop}%
\bibitem [{\citenamefont {Campbell}\ \emph {et~al.}(2010)\citenamefont
  {Campbell}, \citenamefont {Paternostro}, \citenamefont {Bose},\ and\
  \citenamefont {Kim}}]{CampbellPRA2010}%
  \BibitemOpen
  \bibfield  {author} {\bibinfo {author} {\bibfnamefont {S.}~\bibnamefont
  {Campbell}}, \bibinfo {author} {\bibfnamefont {M.}~\bibnamefont
  {Paternostro}}, \bibinfo {author} {\bibfnamefont {S.}~\bibnamefont {Bose}}, \
  and\ \bibinfo {author} {\bibfnamefont {M.~S.}\ \bibnamefont {Kim}},\
  }\bibfield  {title} {\enquote {\bibinfo {title} {Probing the environment of
  an inaccessible system by a qubit ancilla},}\ }\href {\doibase
  10.1103/PhysRevA.81.050301} {\bibfield  {journal} {\bibinfo  {journal} {Phys.
  Rev. A}\ }\textbf {\bibinfo {volume} {81}},\ \bibinfo {pages} {050301}
  (\bibinfo {year} {2010})}\BibitemShut {NoStop}%
\bibitem [{\citenamefont {Campbell}\ \emph {et~al.}(2012)\citenamefont
  {Campbell}, \citenamefont {Smirne}, \citenamefont {Mazzola}, \citenamefont
  {Lo~Gullo}, \citenamefont {Vacchini}, \citenamefont {Busch},\ and\
  \citenamefont {Paternostro}}]{CampbellPRA2012}%
  \BibitemOpen
  \bibfield  {author} {\bibinfo {author} {\bibfnamefont {S.}~\bibnamefont
  {Campbell}}, \bibinfo {author} {\bibfnamefont {A.}~\bibnamefont {Smirne}},
  \bibinfo {author} {\bibfnamefont {L.}~\bibnamefont {Mazzola}}, \bibinfo
  {author} {\bibfnamefont {N.}~\bibnamefont {Lo~Gullo}}, \bibinfo {author}
  {\bibfnamefont {B.}~\bibnamefont {Vacchini}}, \bibinfo {author}
  {\bibfnamefont {Th.}\ \bibnamefont {Busch}}, \ and\ \bibinfo {author}
  {\bibfnamefont {M.}~\bibnamefont {Paternostro}},\ }\bibfield  {title}
  {\enquote {\bibinfo {title} {Critical assessment of two-qubit post-markovian
  master equations},}\ }\href {\doibase 10.1103/PhysRevA.85.032120} {\bibfield
  {journal} {\bibinfo  {journal} {Phys. Rev. A}\ }\textbf {\bibinfo {volume}
  {85}},\ \bibinfo {pages} {032120} (\bibinfo {year} {2012})}\BibitemShut
  {NoStop}%
\bibitem [{\citenamefont {Popovic}\ \emph {et~al.}(2018)\citenamefont
  {Popovic}, \citenamefont {Vacchini},\ and\ \citenamefont
  {Campbell}}]{MariaPRA}%
  \BibitemOpen
  \bibfield  {author} {\bibinfo {author} {\bibfnamefont {M.}~\bibnamefont
  {Popovic}}, \bibinfo {author} {\bibfnamefont {B.}~\bibnamefont {Vacchini}}, \
  and\ \bibinfo {author} {\bibfnamefont {S.}~\bibnamefont {Campbell}},\
  }\bibfield  {title} {\enquote {\bibinfo {title} {Entropy production and
  correlations in a controlled non-markovian setting},}\ }\href {\doibase
  10.1103/PhysRevA.98.012130} {\bibfield  {journal} {\bibinfo  {journal} {Phys.
  Rev. A}\ }\textbf {\bibinfo {volume} {98}},\ \bibinfo {pages} {012130}
  (\bibinfo {year} {2018})}\BibitemShut {NoStop}%
\bibitem [{\citenamefont {Hamedani~Raja}\ \emph {et~al.}(2018)\citenamefont
  {Hamedani~Raja}, \citenamefont {Borrelli}, \citenamefont {Schmidt},
  \citenamefont {Pekola},\ and\ \citenamefont {Maniscalco}}]{SabrinaPRA2018}%
  \BibitemOpen
  \bibfield  {author} {\bibinfo {author} {\bibfnamefont {S.}~\bibnamefont
  {Hamedani~Raja}}, \bibinfo {author} {\bibfnamefont {M.}~\bibnamefont
  {Borrelli}}, \bibinfo {author} {\bibfnamefont {R.}~\bibnamefont {Schmidt}},
  \bibinfo {author} {\bibfnamefont {J.~P.}\ \bibnamefont {Pekola}}, \ and\
  \bibinfo {author} {\bibfnamefont {S.}~\bibnamefont {Maniscalco}},\ }\bibfield
   {title} {\enquote {\bibinfo {title} {Thermodynamic fingerprints of
  non-markovianity in a system of coupled superconducting qubits},}\ }\href
  {\doibase 10.1103/PhysRevA.97.032133} {\bibfield  {journal} {\bibinfo
  {journal} {Phys. Rev. A}\ }\textbf {\bibinfo {volume} {97}},\ \bibinfo
  {pages} {032133} (\bibinfo {year} {2018})}\BibitemShut {NoStop}%
\bibitem [{\citenamefont {Souza}\ \emph {et~al.}(2013)\citenamefont {Souza},
  \citenamefont {Li}, \citenamefont {Soares-Pinto}, \citenamefont {Sarthour},
  \citenamefont {Oliveira}, \citenamefont {Huelga}, \citenamefont
  {Paternostro},\ and\ \citenamefont {Semi\~ao}}]{Souza13}%
  \BibitemOpen
  \bibfield  {author} {\bibinfo {author} {\bibfnamefont {A.~M.}\ \bibnamefont
  {Souza}}, \bibinfo {author} {\bibfnamefont {J.}~\bibnamefont {Li}}, \bibinfo
  {author} {\bibfnamefont {D.~O.}\ \bibnamefont {Soares-Pinto}}, \bibinfo
  {author} {\bibfnamefont {R.~S.}\ \bibnamefont {Sarthour}}, \bibinfo {author}
  {\bibfnamefont {I.~S.}\ \bibnamefont {Oliveira}}, \bibinfo {author}
  {\bibfnamefont {S.~F.}\ \bibnamefont {Huelga}}, \bibinfo {author}
  {\bibfnamefont {M.}~\bibnamefont {Paternostro}}, \ and\ \bibinfo {author}
  {\bibfnamefont {F.~L.}\ \bibnamefont {Semi\~ao}},\ }\bibfield  {title}
  {\enquote {\bibinfo {title} {Experimental demonstration of non-markovian
  dynamics via a temporal bell-like inequality},}\ }\href
  {https://arxiv.org/abs/1308.5761} {\bibfield  {journal} {\bibinfo  {journal}
  {arXiv:1308.5761}\ } (\bibinfo {year} {2013})}\BibitemShut {NoStop}%
\bibitem [{\citenamefont {Rivas}\ \emph {et~al.}(2010)\citenamefont {Rivas},
  \citenamefont {Huelga},\ and\ \citenamefont {Plenio}}]{Rivas2010}%
  \BibitemOpen
  \bibfield  {author} {\bibinfo {author} {\bibfnamefont {A}~\bibnamefont
  {Rivas}}, \bibinfo {author} {\bibfnamefont {S.~F.}\ \bibnamefont {Huelga}}, \
  and\ \bibinfo {author} {\bibfnamefont {M.~B.}\ \bibnamefont {Plenio}},\
  }\href@noop {} {\bibfield  {journal} {\bibinfo  {journal} {Phys. Rev. Lett.}\
  }\textbf {\bibinfo {volume} {105}},\ \bibinfo {pages} {050403} (\bibinfo
  {year} {2010})}\BibitemShut {NoStop}%
\bibitem [{\citenamefont {Yunger~Halpern}\ \emph {et~al.}(2016)\citenamefont
  {Yunger~Halpern}, \citenamefont {Faist}, \citenamefont {Oppenheim},\ and\
  \citenamefont {Winter}}]{Nicole1}%
  \BibitemOpen
  \bibfield  {author} {\bibinfo {author} {\bibfnamefont {N.}~\bibnamefont
  {Yunger~Halpern}}, \bibinfo {author} {\bibfnamefont {P.}~\bibnamefont
  {Faist}}, \bibinfo {author} {\bibfnamefont {J.}~\bibnamefont {Oppenheim}}, \
  and\ \bibinfo {author} {\bibfnamefont {A.}~\bibnamefont {Winter}},\
  }\bibfield  {title} {\enquote {\bibinfo {title} {Noncommuting conserved
  charges in quantum many-body thermalization},}\ }\href {\doibase
  10.1038/ncomms12051} {\bibfield  {journal} {\bibinfo  {journal} {Nat.
  Commun}\ }\textbf {\bibinfo {volume} {7}},\ \bibinfo {pages} {12051}
  (\bibinfo {year} {2016})}\BibitemShut {NoStop}%
\bibitem [{\citenamefont {Yunger~Halpern}\ \emph {et~al.}(2020)\citenamefont
  {Yunger~Halpern}, \citenamefont {Beverland},\ and\ \citenamefont
  {Kalev}}]{Nicole2}%
  \BibitemOpen
  \bibfield  {author} {\bibinfo {author} {\bibfnamefont {N.}~\bibnamefont
  {Yunger~Halpern}}, \bibinfo {author} {\bibfnamefont {M.~E.}\ \bibnamefont
  {Beverland}}, \ and\ \bibinfo {author} {\bibfnamefont {A.}~\bibnamefont
  {Kalev}},\ }\bibfield  {title} {\enquote {\bibinfo {title} {Noncommuting
  conserved charges in quantum many-body thermalization},}\ }\href {\doibase
  10.1103/PhysRevE.101.042117} {\bibfield  {journal} {\bibinfo  {journal}
  {Phys. Rev. E}\ }\textbf {\bibinfo {volume} {101}},\ \bibinfo {pages}
  {042117} (\bibinfo {year} {2020})}\BibitemShut {NoStop}%
\bibitem [{\citenamefont {Arvidsson-Shukur}\ \emph {et~al.}(2021)\citenamefont
  {Arvidsson-Shukur}, \citenamefont {Drori},\ and\ \citenamefont
  {Yunger~Halpern}}]{Nicole3}%
  \BibitemOpen
  \bibfield  {author} {\bibinfo {author} {\bibfnamefont {D.~R.~M.}\
  \bibnamefont {Arvidsson-Shukur}}, \bibinfo {author} {\bibfnamefont {J.~C.}\
  \bibnamefont {Drori}}, \ and\ \bibinfo {author} {\bibfnamefont
  {N.}~\bibnamefont {Yunger~Halpern}},\ }\bibfield  {title} {\enquote {\bibinfo
  {title} {Noncommuting conserved charges in quantum many-body
  thermalization},}\ }\href {\doibase 10.1088/1751-8121/ac0289} {\bibfield
  {journal} {\bibinfo  {journal} {J. Phys. A}\ }\textbf {\bibinfo {volume}
  {54}},\ \bibinfo {pages} {284001} (\bibinfo {year} {2021})}\BibitemShut
  {NoStop}%
\end{thebibliography}%

\end{document}